\documentclass[iop, apjl, tighten]{emulateapj}

\usepackage{apjfonts}
\usepackage{amsmath}
\usepackage{amssymb}
\usepackage{bm}
\usepackage[backref,breaklinks,colorlinks,citecolor=blue]{hyperref}
\usepackage[all]{hypcap}

\usepackage{enumitem}
\usepackage{booktabs}
\usepackage[normalem]{ulem}

\shorttitle{EDGES High-Band Results: III.}
\shortauthors{Monsalve et al.}

\begin{document}

\title{Results from EDGES High-Band: III. New Constraints on Parameters of the Early Universe}

\author{
Raul A. Monsalve\altaffilmark{1,2,3,4,5},
Anastasia Fialkov\altaffilmark{6,7,8,9,10},
Judd D. Bowman\altaffilmark{3}, 
Alan E. E. Rogers\altaffilmark{11},
Thomas J. Mozdzen\altaffilmark{3},
Aviad Cohen\altaffilmark{12},
Rennan Barkana\altaffilmark{12},
and Nivedita Mahesh\altaffilmark{3}
}

\affil{$^1$Department of Physics, McGill University, Montr\'eal, 3600 Rue University, QC H3A 2T8, Canada; \href{mailto:Raul.Monsalve@physics.mcgill.ca}{Raul.Monsalve@mcgill.ca}}
\affil{$^2$McGill Space Institute, McGill University, Montr\'eal, 3550 Rue University, QC H3A 2A7, Canada}
\affil{$^3$School of Earth and Space Exploration, Arizona State University, 781 Terrace Mall, Tempe, AZ 85287, USA}
\affil{$^4$Center for Astrophysics and Space Astronomy, University of Colorado, 2000 Colorado Avenue, Boulder, CO 80309, USA}
\affil{$^5$Facultad de Ingenier\'ia, Universidad Cat\'olica de la Sant\'isima Concepci\'on, Alonso de Ribera 2850, Concepci\'on, Chile}
\affil{$^6$Harvard-Smithsonian Center for Astrophysics, Institute for Theory and Computation, 60 Garden Street, Cambridge, MA 02138, USA} % 
\affil{$^7$Institute of Astronomy, University of Cambridge, Madingley Road, Cambridge CB3 0HA, UK}
\affil{$^8$Kavli Institute for Cosmology, University of Cambridge, Madingley Road, Cambridge CB3 0HA, UK}
\affil{$^9$Department of Physics, The University of Tokyo, 7-3-1 Hongo, Bunkyo, Tokyo 113-0033, Japan}
\affil{$^{10}$Department of Physics and Astronomy, University of Sussex, Falmer, Brighton BN1 9QH, UK}
\affil{$^{11}$Haystack Observatory, Massachusetts Institute of Technology, 99 Millstone Road, Westford, MA 01886, USA}
\affil{$^{12}$Raymond and Beverly Sackler School of Physics and Astronomy, Tel Aviv University, Tel Aviv 69978, Israel}

% Obtaining independent constraints on astrophysical parameters of the early Universe remains a critical objective in cosmology after the report of an absorption feature in the radio spectrum by EDGES Low-Band. Here 

% {\color{red} ASK PARKS WHY DIDN'T MENTION EDGES CONSTRAINTS. THEN, READ THE WHOLE THING ONCE AGAIN. CHECK ALL TEXT, INCLUDING ABSTRACT, AND DECIDE IF RESULTS SHOULD BE GIVEN AS DISFAVORED REGIONS, OR PREFERRED REGIONS} 

\begin{abstract}
We present new constraints on parameters of cosmic dawn and the epoch of reionization derived from the EDGES High-Band spectrum ($90-190$~MHz). The parameters are probed by evaluating global $21$~cm signals generated with the recently developed \texttt{Global21cm} tool. This tool uses neural networks trained and tested on $\sim 30,000$ spectra produced with semi-numerical simulations that assume the standard thermal evolution of the cosmic microwave background and the intergalactic medium. From our analysis, we constrain at $68\%$ (1) the minimum virial circular velocity of star-forming halos to $V_{\rm c}<19.3$~km~s$^{-1}$, (2) the X-ray heating efficiency of early sources to $f_{\rm X}>0.0042$, and (3) the low-energy cutoff of the X-ray spectral energy distribution to $\nu_{\rm min}<2.3$~keV. We also constrain the star-formation efficiency ($f_*$), the electron scattering optical depth ($\tau_{\rm e}$), and the mean-free path of ionizing photons ($R_{\rm mfp}$). We re-compute the constraints after incorporating into the analysis four estimates for the neutral hydrogen fraction from high-$z$ quasars and galaxies, and a prior on $\tau_{\rm e}$ from \emph{Planck}~$2018$. The largest impact of the external observations is on the parameters that most directly characterize reionization. Specifically, we derive the combined $68\%$ constraints $\tau_{\rm e}<0.063$ and $R_{\rm mfp}>27.5$~Mpc. The external observations also have a significant effect on $V_{\rm c}$ due to its degeneracy with $\tau_{\rm e}$, while the constraints on $f_*$, $f_{\rm X}$, and $\nu_{\rm min}$, remain primarily determined by EDGES.
\end{abstract}


% The external observations also reduce significantly the probability of low $R_{\rm mfp}$; the combined 

% ; combining EDGES with the external observations disfavors $V_{\rm c}<6.0$ and $V_{\rm c}>46.6$~km~s$^{-1}$ ($68\%$). The external observations also prefer late and fast reionization, which drives the constraint on $R_{\rm mfp}$; specifically, combining all the observations yields the lower limit $R_{\rm mfp}>27.5$~Mpc ($68\%$), reversing the result obtained from EDGES alone. In the combined analysis, 

\section{Introduction}
\label{section_introduction}

The sky-averaged, or global, radio spectrum is expected to encode the redshift evolution of the $21$~cm line of neutral hydrogen gas in the intergalactic medium (IGM) during the formation of the first stars and galaxies \citep{varshalovich1977, tozzi2000, barkana2001, furlanetto2006, pritchard2008}. At the onset of star formation, stellar Ly-$\alpha$ photons couple the spin temperature of the $21$~cm signal to the temperature of the gas, a process that makes the line visible in absorption relative to the radio background radiation \citep{wouthuysen1952, field1958}. The detection by \citet{bowman2018} of an absorption feature at $\sim78$ MHz, with a width of $\sim20$~MHz and an amplitude of $\sim0.5$~K, if confirmed to be of cosmological origin, is the first measurement of the $21$~cm signal from cosmic dawn and would represent direct evidence of the formation of the first stars. The central frequency of the reported feature is in agreement with theoretical predictions and implies efficient star formation in halos of mass below $10^8$ M$_{\odot}$ \citep{mirocha2019}. Such a population would manifest itself in future high-redshift galaxy surveys generating a steeper than expected UV luminosity function at the faint end \citep{mirocha2017, mirocha2019}.

However, the amplitude and shape of the detected radio signal do not comply with standard models of cosmic dawn and reionization, where the absorption is measured against the cosmic microwave background (CMB) and the IGM is heated by X-ray sources after an initial period of adiabatic cooling \citep[e.g,][]{mesinger2011, cohen2017, mirocha2018}. The reported absorption amplitude is at least twice as large as predicted  and the observed shape is much flatter at the bottom than expected. These discrepancies have led to many suggestions of exotic physical mechanisms that could produce either a colder IGM temperature at early times \citep[e.g.,][]{tashiro2014, munoz2015, barkana2018, munoz2018, fialkov2018, barkana2018b, berlin2018, hektor2018, sikivie2018, houston2018} or a radiation background stronger than the CMB \citep[e.g.,][]{feng2018, ewallwice2018, fraser2018}. 

Verification of the unexpected EDGES measurement, which was conducted with two `Low-Band' instruments observing in the $50-100$ MHz frequency range ($26.4 \gtrsim z \gtrsim 13.2$), requires independent observations and constraints. Apart from the EDGES result, the only existing constraint on the global $21$~cm signal from cosmic dawn is the upper limit for the absorption amplitude of $0.89$~K ($95\%$), in the same frequency range, established by the LEDA experiment \citep{bernardi2016}. Previously, the SCI-HI experiment reported upper limits in the range $\sim 1-10$~K on the residual spectral structure after removing a model for the foreground contribution \citep{voytek2014}. Upper limits on the $21$~cm power spectrum signal have been presented in the range $z=12-18$ by MWA \citep{ewallwice2016} and $z=20-25$ by LOFAR \citep{gehlot2018}. An additional, but also currently disputed, potential evidence of new physics at cosmic dawn could come from the measurement of the brightness temperature of the diffuse sky by ARCADE~2, which found a $54$-mK `excess' above the CMB at $3.3$~GHz \citep{fixen2011}. A consistent excess was recently reported using LWA1 data over $40-80$~MHz \citep{dowell2018}. However, the existence of this excess relies on the correct identification and removal of the Galactic and extragalactic contributions. Using a more realistic Galactic model, \citet{subrahmanyan2013} showed that the excess could instead correspond to a Galactic contribution not accounted for in other analyses.
 
Compared to cosmic dawn, constraints on the epoch of reionization are tighter and stem from several independent observational probes, none of which has so far reported deviations from traditional astrophysics.  The average fraction of neutral hydrogen in the IGM, $\bar{x}_{\rm H\scriptscriptstyle{I}}$, can be constrained through the Ly-$\alpha$ and Ly-$\beta$ pixels that are dark in the spectra of high-$z$ quasars. Using this technique and a sample of $22$ quasars at $z = 5-6$, \citet{mcgreer2015} derive the upper limit $\bar{x}_{\rm H\scriptscriptstyle{I}}\leq 0.06+0.05\;(68\%)$ at $z = 5.9$. Neutral hydrogen in the IGM also imprints a `damping wing' absorption feature in the spectrum of high-$z$ quasars. \citet{greig2017c} obtain $\bar{x}_{\rm H\scriptscriptstyle{I}} = 0.40^{+0.21 }_{-0.19}$ ($68\%$) from the damping wing analysis of ULASJ1120+0641, showing that reionization is ongoing at $z=7.08$; while the spectrum of ULASJ1342+0928, the highest-redshift quasar detected so far, yields $\bar{x}_{\rm H\scriptscriptstyle{I}} = 0.65^{+0.15 }_{-0.32}$ ($68\%$) at $z=7.54$ in a damping wing analysis by \citet{banados2018}. The IGM neutral fraction can also be constrained from the emission of Ly-$\alpha$ radiation from Lyman Break galaxies (LBGs). In a Bayesian analysis that incorporates reionization simulations and empirical models of the interstellar medium, \citet{mason2018} determine $\bar{x}_{\rm H\scriptscriptstyle{I}} = 0.59^{+0.11 }_{-0.15}$ ($68\%$) at $z=7$ from a sample of LBGs presented in \citet{pentericci2014}. CMB fluctuations provide an independent test of reionization by probing the integrated electron scattering optical depth to recombination, $\tau_{\rm e}$. Among other values, the \emph{Planck} satellite recently reported $\tau_{\rm e}=0.056\pm 0.007$, which corresponds to a reionization center redshift $z=7.82\pm0.71$ assuming a `tanh' phenomenological model \citep{planck2018}. For a similar type of model, the data from the EDGES `High-Band' instrument provide independent constraints on the reionization duration ($\Delta z$) via non-detection of the $21$~cm line in the $90-190$ MHz range \citep{monsalve2017b}: at $\geq 2\sigma$ significance, the data rule out models with $\Delta z < 1$ at $z \sim 8.5$ and higher than $\Delta z=0.4$ across most of the $14.8 \geq z \geq 6.5$ range. These constraints implicitly correspond to reionization scenarios where the $21$~cm spin temperature of neutral hydrogen is saturated, i.e. much higher than the microwave background, due to prior IGM heating. \citet{monsalve2017b} also explored and ruled out $21$~cm models that take the opposite extreme assumption, i.e., total Ly-$\alpha$ coupling but no IGM heating before reionization. In this case, the hydrogen neutral fraction was also modeled using the `tanh' expression. As a reference result, \citet{monsalve2017b} ruled out at $\geq 2\sigma$ all the reionization models with total Ly-$\alpha$ coupling but no IGM heating that produce $\bar{x}_{\rm H\scriptscriptstyle{I}} \leq 1\%$ at $z=6$ and have an optical depth in the range $0.086\geq \tau_{\rm e} \geq 0.038$. Finally, there are also constraints on the $21$~cm power-spectrum signal at $z=8.6$ from GMRT \citep{paciga2013} and at $z=7.1$ from MWA \citep{beardsley2016}, while the tightest upper limits have been reported by LOFAR in the range $z=9.6-10.6$ \citep{patil2017}\footnote{The PAPER $21$~cm power spectrum constraint at $z=8.4$ of \citet{ali2015} has been retracted in \citet{ali2018}.}.

Recently, global radio spectra were analyzed using astrophysical models for the first time  \citep{singh2017, singh2018, monsalve2018}. A set of $193$ models from a parameter study by \citet{cohen2017} was  evaluated using data from the SARAS~2 experiment in the $110-200$ MHz band, which allowed to rule out, at $>5\sigma$ significance, $25$  models that share inefficient X-ray heating and rapid reionization \citep{singh2017, singh2018}. A much broader study was done using EDGES High-Band data \citep[][M18 hereafter]{monsalve2018}, which evaluated $10,000$ models generated with the \texttt{21cmFAST} code \citep{mesinger2007, mesinger2011}. M18 constrained the following parameters of cosmic dawn and reionization: the minimum virial temperature ($T_{\rm vir}^{\rm min}$) and ionizing efficiency ($\zeta$) of star-forming halos, as well as the integrated soft-band X-ray luminosity ($L_{\rm{X<2 keV}}/{\rm SFR}$) and threshold energy for self-absorption ($E_0$) of the first galaxies. For reference, using EDGES data alone M18 disfavored ($68\%$) $\log_{10}(T_{\rm vir}^{\rm min}/{\rm K})>5.5$ and $\zeta>154.6$, as well as the intermediate range of X-ray luminosity $38.8<\log_{10}\left(L_{\rm{X<2 keV}}/{\rm SFR}\;{\rm /erg\;yr\;s}^{-1}\;{\rm M}^{-1}_{\odot}\right)<40.4$. Further, combining (1) the EDGES High-Band data, (2) an estimate for $\tau_{\rm e}$ from \emph{Planck}~$2016$, and (3) constraints on $\bar{x}_{\rm H\scriptscriptstyle{I}}$ from quasars at $z=5.9$ and $z=7.08$, resulted in significantly stronger constraints on $\zeta$ and $T_{\rm vir}^{\rm min}$, with EDGES contributing to produce slightly better results than those derived in \citet{greig2017a} using only the information on $\tau_{\rm e}$ and $\bar{x}_{\rm H\scriptscriptstyle{I}}$.

In this paper we use the EDGES High-Band data to evaluate a different set of astrophysical models, which were generated with the \texttt{Global21cm} global signal emulator described in detail by \citet{cohen2019}. This tool is based on neural networks trained on $29,641$ outputs of semi-numerical simulations of cosmic dawn and reionization described in detail in \citet{visbal2012, fialkov2014, cohen2017}. The simulations make standard assumptions for the temperatures of the CMB and the IGM, not taking into account the exotic physics invoked to explain the EDGES Low-Band result. To produce the simulations, seven astrophysical parameters were varied in the widest possible range: the minimum virial circular velocity of star-forming halos ($V_{\rm c}$), the star formation efficiency ($f_*$), the X-ray heating efficiency of early sources ($f_{\rm X}$), the low-energy cutoff ($\nu_{\rm min}$) of the X-ray spectral energy distribution (SED), the slope ($\alpha$) of the X-ray SED, the mean-free path of ionizing photons ($R_{\rm mfp}$), and $\tau_{\rm e}$. \texttt{Global21cm} interpolates between the outputs of the semi-numerical simulations and produces global signals for any combination of parameters.

Here, we generate $6.4$ million global signals using \texttt{Global21cm} and conduct a Bayesian analysis that rigorously maps the posterior probability density function (PDF) of six of the astrophysical parameters: $V_{\rm c}$, $f_*$, $f_{\rm X}$, $\nu_{\rm min}$, $R_{\rm mfp}$, and $\tau_{\rm e}$. We fix the value of $\alpha$ as it only mildly affects the results. We derive constraints on the parameters first using the EDGES High-Band data alone, and then combining them with a prior on $\tau_{\rm e}$ from \emph{Planck} and constraints on $\bar{x}_{\rm H\scriptscriptstyle{I}}$ from high-$z$ quasars and galaxies. Our main results are the 1D and 2D PDFs of each parameter and parameter pair, obtained after marginalizing over the rest of the astrophysical parameters as well as the parameters that account for the contribution of diffuse foregrounds to the radio spectrum.

In addition to exploring more parameters --- six instead of four ---, we improve on M18 by including in our combined analysis an updated prior on $\tau_{\rm e}$ (\emph{Planck}~$2018$ instead of $2016$), a constraint on $\bar{x}_{\rm H\scriptscriptstyle{I}}$ at $z=7.54$ by \citet{banados2018} from the ULASJ1342+0928 quasar, and a constraint on $\bar{x}_{\rm H\scriptscriptstyle{I}}$ at $z=7$ by \citet{mason2018} from LBGs. In general, the parameters we explore are different from those in M18 and, although some of them overlap, here we explore them over a wider range. This makes it difficult to compare in detail our results with M18. On the other hand, it enables us to derive independent conclusions about the astrophysics of the early Universe. As in M18, we do not incorporate the EDGES Low-Band spectrum into our analysis, saving that for future work.

The paper is organized as follows. In Section~\ref{section_parameters} we briefly describe the $21$~cm astrophysical parameters. In Section~\ref{section_analysis} we detail our analysis procedure. In Section~\ref{section_results} we present the results obtained from the analysis of EDGES data alone, the external constraints alone, and their combination. In Section~\ref{section_discussion} we discuss the results and compare them with those for models from \texttt{21cmFAST}. Finally, in Section~\ref{section_summary} we summarize this work.

\section{Astrophysical Parameters}
\label{section_parameters}

The \texttt{Global21cm} code outputs in less than a second a global $21$~cm signal over the redshift range $6<z<50$ given a combination of key astrophysical parameters. The code employs neural networks that were trained on $29,641$ global spectra produced with a hybrid simulation of the high-redshift Universe  \citep{visbal2012,fialkov2014,cohen2017}. For an input set of astrophysical parameters, the simulation  generates a realization of the $21$~cm signal within large cosmological volumes ($384^3$ co-moving Mpc$^3$) and over a wide redshift range ($z=6-60$). The global spectra are obtained by averaging the three-dimensional $21$~cm fields over the box at every redshift. Each simulation takes $\sim 4$ hours to run on a desktop, and the ensemble of $29,641$ models was produced using the {\it Odyssey} cluster at Harvard University\footnote{\url{https://www.rc.fas.harvard.edu/odyssey/}}.  All these runs were executed with the same set of initial conditions for large-scale density and velocity fields at $z=60$, and assume $\Lambda$CDM with the standard cosmological parameters \citep{planck2014}. The simulation follows the hierarchical growth of structure, tracks star formation averaged over scales of $\sim3$~Mpc, and follows the evolution of inhomogeneous Ly-$\alpha$, Lyman-Werner ($11.2-13.6$~eV), X-ray, and ionizing radiative backgrounds. The simulation takes into account the effect of relative streaming velocity \citep{visbal2012}, Lyman-Werner radiation \citep{fialkov2013}, and photoheating feedback on star formation \citep{cohen2016}.

In the simulation, the high-redshift astrophysics is parametrized with seven parameters: $ V_{\rm c}$, $f_*$, $f_{\rm X}$, $\nu_{\rm min}$, $\alpha$, $R_{\rm mfp}$, and $\zeta$. The \texttt{Global21cm} code receives as input the same parameters except for $\zeta$; the code uses instead the CMB optical depth, which is related to $\zeta$ by a one-to-one mapping. In the rest of this section we briefly describe the parameters and their impact on the global signal. For detailed descriptions we point the reader to \citet{cohen2017, cohen2019}. An example is given in Figure \ref{figure_signals}, where we show the effect of changes in all the parameters except $\alpha$, which is kept fixed at $\alpha=-1.3$. In Table~\ref{table_ranges} we list the parameter ranges explored in this paper, as well as the scale used to sample these ranges with \texttt{Global21cm}. Although our parametrization intends to characterize the large-scale physics of cosmic dawn and reionization, it is not necessarily optimized for the global $21$~cm signal. Therefore, the effect of some of the parameters on the global signal is degenerate. We discuss this point in Section~\ref{section_degeneracy}.

\capstartfalse
\begin{deluxetable}{lrrll}
%\tabletypesize{\scriptsize}
\tablewidth{0pt}
\tablecaption{Parameter ranges and sampling scale \label{table_ranges}   }
\tablehead{Parameter & Min & Max & Unit & Scale}
\startdata
$V_{\rm c}$         \dotfill     & $4.2$      & $76.5$   & km\;$\rm s^{-1}$ & $\log_{10}$ \\
$f_*$               \dotfill     & $10^{-3}$  & $0.5$    &                  & $\log_{10}$ \\
$f_{\rm X}$         \dotfill     & $10^{-5}$  & $10$     &                  & $\log_{10}$ \\
$\nu_{\rm min}$     \dotfill     & $0.1$      & $3$      & keV              & linear \\
$\tau_{\rm e}$      \dotfill     & $0.055$    & $0.09$   &                  & linear \\
$R_{\rm {mfp}}$     \dotfill     & $10$       & $50$     & Mpc              & linear
\enddata
\end{deluxetable}

\begin{itemize}[leftmargin=*, label={}]
\item \emph{Minimum Virial Circular Velocity}: In the hierarchical picture of structure formation, low-mass halos form at higher redshifts and are more numerous than high-mass halos. Therefore, in the cases with lower threshold mass for star formation, $M_{\rm min}$, stars form earlier, leading to an earlier onset of the Ly-$\alpha$ coupling and shifting the descending slope of the $21$~cm absorption feature to lower frequencies. In \texttt{Global21cm} we use the minimum virial circular velocity,

\begin{equation}
V_{\rm c} \sim 16.9 \left(\frac{M_{\rm min}}{10^8} \right)^{1/3}\left(\frac{1+z}{10}\right)^{1/2}~{\rm km~ s}^{-1},
\end{equation}

\noindent instead of $M_{\rm min}$ because $V_{\rm c}$ is less dependent on redshift \citep{barkana2016}.

$V_{\rm c}$ is primarily set  by the cooling channel. Molecular hydrogen cooling fuels star formation in halos with $V_{\rm c}=4.2-16.5$~km~s$^{-1}$, which corresponds to halo masses $M_{\rm h} \sim 1\times 10^{6}-8\times 10^{7}$~M$_\odot$ at $z=10$, while atomic hydrogen cooling occurs for $V_{\rm c}>16.5$~km~s$^{-1}$ \citep{barkana2016}. In addition to radiative cooling, star formation is affected by numerous feedback processes that result in spatial variation of  $V_{\rm c}$ as well as in its dependence on background radiation fields \citep[e.g.,][]{machacek2001, fialkov2012, fialkov2013}. To probe different cooling and feedback mechanisms, in this paper we vary $V_{\rm c}$ in a broad range, from $4.2$~km~s$^{-1}$, corresponding to the minimum value for molecular hydrogen cooling, to $76.5$~km~s$^{-1}$, representing atomic hydrogen cooling and inefficient star formation in smaller halos due to, e.g., supernovae feedback.

\item \emph{Star Formation Efficiency}: The star formation efficiency (SFE) corresponds to the fraction of gas in dark matter halos that is converted into stars.  Higher values of SFE result in an earlier onset of Ly-$\alpha$ coupling, as well as in a faster build-up of X-ray and ionizing radiation backgrounds. The SFE depends on feedback mechanisms, the metalicity of the gas, and the halo mass. Star formation simulations in metal-poor environments show large scatter, with SFE between $\sim0.1\%$ and $\sim10\%$ in halos of $M_{\rm h} \sim 10^8-10^{10}$~M$_\odot$ at $z\sim 10$  \citep[e.g.,][]{xu2016, ceverino2017, ma2018}. Abundance matching techniques applied to $z\geq 6$ galaxies find that the SFE peaks at $\sim 30$\% for halos of $M_{\rm h} \sim 10^{11} - 10^{12}$~M$_\odot$ dropping to $\sim 10\%$ at $M_{\rm h} \sim 10^{10}$~M$_\odot$ and $M_{\rm h} \sim 10^{13}$~M$_\odot$ \citep{behroozi2015, mason2015, mashian2016, sun2016, mirocha2017}. Finally, observations of dwarf galaxies infer SFE~$\sim 0.01 - 0.1 \%$ at $z\sim10$ \citep{read2016}.

We use the following SFE-$M_{\rm h}$ dependence for the models in this study \citep{cohen2017}:

\begin{equation}
\text{SFE}(M_{\rm h}) = \left\{ \begin{array}{ll}
f_* & M_{\rm atomic} < M_{\rm h} \ ,\\
f_*\frac{\log\left( M_{\rm h}/M_{\rm min}\right) }{\log\left( M_{\rm atomic}/M_{\rm min}\right) } & M_{\rm min} < M_{\rm h} < M_{\rm atomic} \ ,\\
0 & M_{\rm h} < M_{\rm min}\ ,\end{array} \right.
\label{equation_fstar}
\end{equation}

\noindent 
where $M_{\rm min}$ is the minimum cooling mass of star-forming halos, $M_{\rm atomic}$ is the minimum halo mass for atomic cooling, and $f_*$ is a parameter that stands for the SFE at the high-mass end. We vary $f_*$ over the wide range $0.1\%-50\%$.

\item \emph{X-ray SED Low-frequency Cutoff}: The process of IGM X-ray heating can be characterized in terms of the shape of the X-ray SED and the total luminosity. The IGM heating rate, and thus the evolution of the gas temperature, depends on the amount of energy injected by X-ray sources below $\sim2$~keV. Dust in host galaxies prevents soft X-rays below $\mathcal{O}(0.1)$~keV from penetrating into the IGM, imposing a low-energy cutoff in the spectrum of the injected photons \citep[e.g.,][]{das2017}. On the other hand, hard X-rays are barely absorbed; they free-stream and add up to form an X-ray background. 

In \texttt{Global21cm} we model the X-ray SED as a power-law with slope $\alpha$ and low-energy cutoff $\nu_{\rm min}$. However, we find that variations in $\alpha$ lead to very weak variations of the global signal relative to the sensitivity of the EDGES High-Band data. Therefore, here we fix the value of $\alpha$ at $-1.3$. Higher values of  $\nu_{\rm {min}}$ lead to the effective hardening of the X-ray SED, less efficient heating, and, as a result, deeper $21$~cm absorption with a higher central frequency. This scenario resembles the effect of X-ray binaries (XRBs), one of the most plausible sources to dominate high-redshift X-ray emission. XRBs are expected to have a hard X-ray SED that peaks at $\sim1-3$ keV and has a high-energy tail following a power-law with slope $\alpha \sim -1.5$ \citep{mirabel2011, fragos2013}. We vary $\nu_{\rm {min}}$ in the range $0.1-3$~keV, which is wide enough to explore the effects of host galaxy absorption as well as hard X-rays.

\item \emph{X-ray Efficiency}: The total X-ray luminosity of early sources satisfies the following relation, derived from observations of nearby starburst galaxies and XRBs \citep{grimm2003, gilfanov2004, mineo2012}:

\begin{equation}
\frac{L_{\rm X}}{\rm SFR}=3\times10^{40}f_{\rm X}\;\;\;\;\;\;\;\rm~erg~s^{-1}~M_\odot^{-1}~yr\ ,
\label{equation_xray}
\end{equation}

\noindent where $L_{\text{X}}$ is the total X-ray luminosity emitted in the range $\nu_{\rm min}-95$~keV, SFR is the star-formation rate (which in our parametrization is a function of $M_{\rm h}$, $f_*$, and $z$, as well as of the large-scale overdensity and relative velocity between dark matter and gas), and $f_{\rm X}$ is the X-ray efficiency of sources, which is our parameter in \texttt{Global21cm}. \citet{fialkov2017} found that the unresolved soft X-ray background measured by the {\it Chandra} X-ray observatory \citep{lehmer2012} imposes an upper limit on $f_{\rm X}$ in the range $\sim10-500$, depending on the nature of the X-ray sources, the halo cooling channel, and the reionization history. For high values of $f_{\rm X}$, the contribution of X-rays to reionization becomes significant \citep[up to $\sim 50\%$ in the case with  $f_{\rm X} = 422$ and $\nu_{\rm min} = 0.2$ keV,][]{fialkov2017}, and the absorption trough is shallow and occurs at low frequencies. Low values of $f_{\rm X}$ result in deep absorption troughs centered at high frequencies. This has enabled to exclude models with low $f_{\rm X}$ (for some values of $V_{\rm c}$ and $f_*$) using SARAS~2 data \citep{singh2017, singh2018}. Here we vary $f_{\rm X}$ over the wide range $10^{-5}-10$.

\item \emph{Mean-free Path of Ionizing Photons}: During reionization, the distance ionizing photons can propagate into the IGM determines the physical size of ionized regions. This distance depends on the abundance, density, and structure of photon sinks --- absorption systems such as Lyman limit systems ---, and the corresponding recombinations of these systems. In our parametrization we explore the mean-free path of ionizing photons, $R_{\rm {mfp}}$, which we vary over $10-50$~Mpc \citep{alvarez2012, greig2017b}. The effect of this parameter is only manifested after the onset of reionization. Higher values of $R_{\rm {mfp}}$ lead to a faster reionization and a steeper $21$~cm signal at the high-frequency end, which can be constrained by EDGES High-Band.

\begin{figure}
\begin{center}
\includegraphics[width=0.48\textwidth]{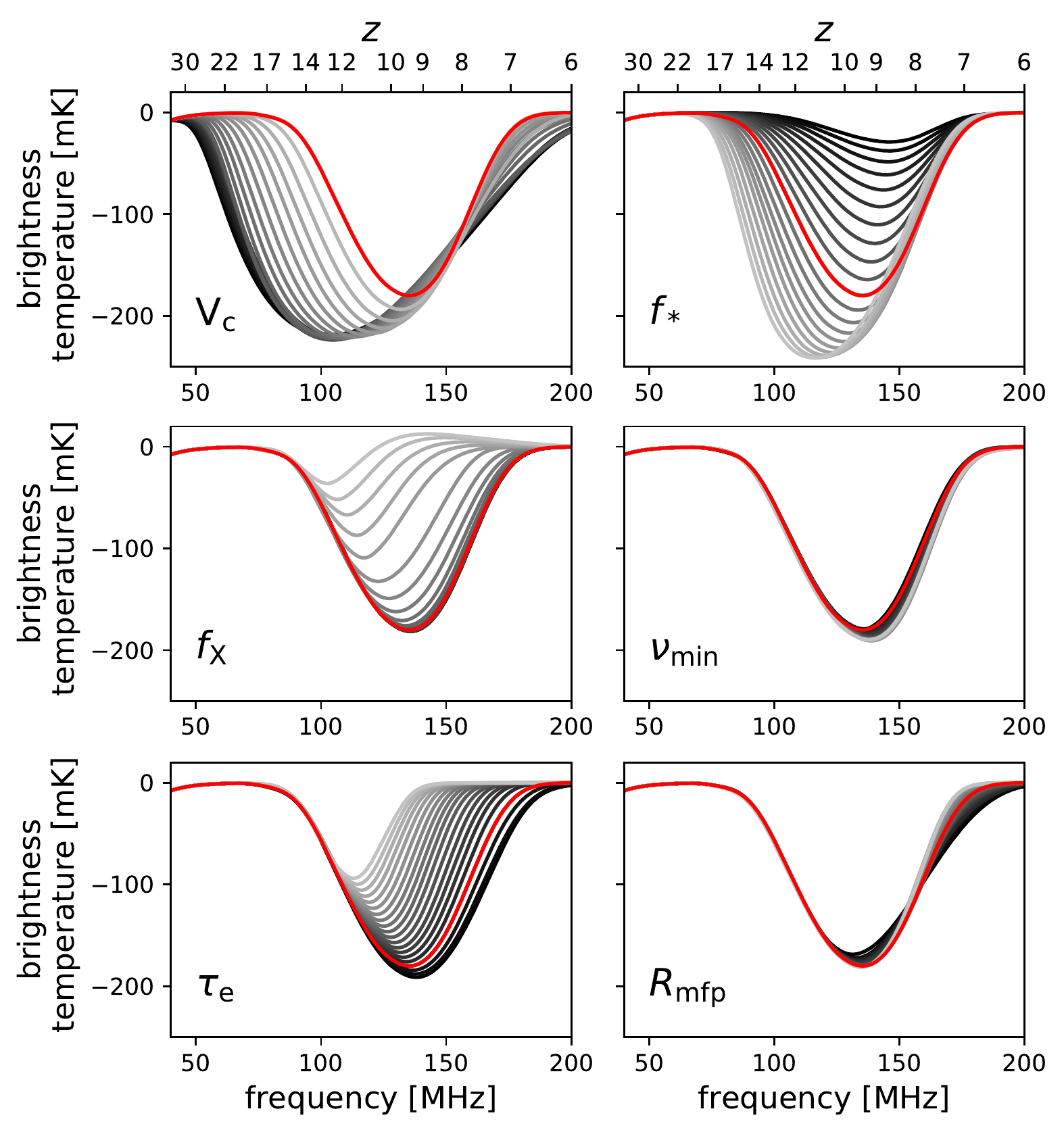}
\caption{Example of dependence of the global $21$~cm signal on the parameters constrained. The signals are produced using the \texttt{Global21cm} code \citep{cohen2019}. Each panel shows variations in one parameter relative to a common reference. The colors of the signals, from black to light gray, represent the parameter value going from lowest to highest, spanning the ranges described in Section \ref{section_parameters} and listed in Table~\ref{table_ranges}. For $V_{\rm c}$, $f_*$, and $f_{\rm X}$, the sampling is done evenly in $\log_{10}$ scale. The red signal is the common reference, with parameters $V_{\rm c}=76.5$~km s$^{-1}$, $f_*=0.026$, $f_{\text{X}}=3.4\times10^{-3}$, $\nu_{\text{min}}=0.25$~keV, $\tau_{\rm e}=0.06$, and $R_{\text{mfp}}=31$~Mpc.}
\label{figure_signals}
\end{center}
\end{figure}

\item \emph{Electron Scattering Optical Depth}: The last independent parameter is the ionizing efficiency of sources, $\zeta$ \citep[][M18]{greig2017a}. However, because the CMB optical depth --- rather than $\zeta$ --- is directly probed by the CMB experiments, \texttt{Global21cm} was constructed to receive $\tau_{\rm e}$ instead of $\zeta$ as an input parameter. The CMB optical depth measures the total column density of ionized gas, and, thus, is a function of the reionization history, $\bar{x}_{\rm H\scriptscriptstyle{I}}$, which is inferred from the simulations and depends on all the astrophysical parameters.  Given the evolution of $\bar{x}_{\rm H\scriptscriptstyle{I}}$ with redshift and for a mass-independent ionizing efficiency, we find a one-to-one relation between $\zeta$ and $\tau_{\rm e}$. The mapping between these two parameters is done using a neural network that was trained on a set of $27,455$ cases and tested with $2,186$ cases \citep{cohen2019}. Increasing $\tau_{\rm e}$ while keeping the other parameters fixed amounts to a higher $\zeta$ and a faster depletion of neutral gas. This results in an earlier reionization and a shallower $21$~cm absorption with the trough shifted to lower frequencies, as well as a reduced emission feature if such exists. Sixty-eight-percent constraints from the \emph{Planck}~2016 release include $\tau_{\rm e}=0.066\pm 0.016$, $0.078\pm 0.019$ \citep{planck2016a}, $0.055\pm 0.009$ \citep{planck2016b}, and $0.058\pm 0.012$ \citep{planck2016c}. Considering these estimates, we explore the range $\tau_{\rm e}=0.055-0.09$. The lower limit of the range, $\tau_{\rm e}=0.055$, was determined from the upper limit on $\bar{x}_{\rm H\scriptscriptstyle{I}}$ reported by \citet{mcgreer2015}, of $\bar{x}_{\rm H\scriptscriptstyle{I}}\leq 0.06+0.05\;(68\%)$ at $z=5.9$. For our \texttt{Global21cm} models, this upper limit results in a $<1\%$ probability for $\tau_{\rm e}<0.056$ when considering the range $\tau_{\rm e}=0.055-0.09$ (see Section~\ref{section_results_combined}). In Section~\ref{section_results_external} we discuss how new $2018$ results from high-$z$ galaxies \citep{mason2018} and \emph{Planck} \citep{planck2018} warrant extending the range to lower values of $\tau_{\rm e}$ in future studies. The upper limit of our range, $\tau_{\rm e}=0.09$, is high considering current constraints. However, exploring a wide range is useful since $\tau_{\rm e}$ is model-dependent and our $\bar{x}_{\rm H\scriptscriptstyle{I}}$ models do not correspond to the ones used by \emph{Planck}.

% , and $0.056\pm 0.007$ \citep{planck2018}

\end{itemize}

% THIS TEXT IS REDUNDANT:
% Over the redshift range that the \texttt{Global21cm} models span ($30>z>6$), the parameters described above affect the global signal more strongly during the development of different physical processes. $V_{\rm c}$ and $f_*$ affect the signal from the onset of star formation through the end of reionization. The X-ray heating parameters, $f_{\rm X}$ and $\nu_{\rm min}$, impact the signal from the moment of formation of the first X-ray sources ($z\sim 20$). Finally, the parameters of reionization, $\tau_{\rm e}$ and $R_{\rm {mfp}}$, only affect the high-frequency end of the signal, with $\tau_{\rm e}$ producing much stronger changes than $R_{\rm {mfp}}$. Overall, the lowest sensitivity of the signal is to changes in $\nu_{\rm min}$ and $R_{\rm {mfp}}$. As we show in this paper, the EDGES High-Band data are sensitive to each one of these parameters via their effect on the signal in the range $14.8>z>6.5$.

\section{Analysis}
\label{section_analysis}

Exploring rigorously the six-dimensional parameter space described in Section \ref{section_parameters} with high resolution is computationally expensive. Considering the low sensitivity of the global $21$~cm signal to changes in $\nu_{\rm min}$ and $R_{\rm {mfp}}$, we explore the six parameters by dividing the space into two subsets of five parameters each. In one subset, the fifth parameter is $\nu_{\rm min}$ and $R_{\rm {mfp}}$ is fixed at $30$~Mpc. In the other subset, the fifth parameter is $R_{\rm {mfp}}$ and $\nu_{\rm min}$ is fixed at $0.5$~keV. We generate the $21$~cm signals by evaluating the \texttt{Global21cm} code at $20$ values per parameter over a regular grid in the ranges described in Section~\ref{section_parameters} and Table~\ref{table_ranges}. This produces a total of $20^5=3.2$ million models for each five-parameter set. Because of their large dynamic ranges, the sampling for  $V_{\rm c}$, $f_*$, and $f_{\rm X}$ is done in $\log_{10}$ scale, while $\tau_{\rm e}$, $\nu_{\rm min}$, and $R_{\rm mfp}$ are sampled in linear scale. 

Following M18, we constrain the parameters by computing their marginalized posterior PDFs within a Bayesian framework. We first derive constraints using EDGES data alone, and then incorporating into the analysis external estimates for $\tau_{\rm e}$ and $\bar{x}_{\rm H\scriptscriptstyle{I}}$. We describe the analyses next.

\subsection{EDGES-only Analysis}

In the EDGES-only analysis we start by fitting our model for the diffuse foregrounds to the difference $d-m_{21}(\bm{\theta}_{21})$, where $d$ is the spectrum measured by EDGES in the range $90-190$~MHz \citep{monsalve2017b} and $m_{21}(\bm{\theta}_{21})$ represents each $21$~cm signal produced evaluating \texttt{Global21cm} at the vector of $21$~cm astrophysical parameters $\bm{\theta}_{21}$. The diffuse foreground model is given by \citep[][M18]{mozden2016, monsalve2017a, monsalve2017b}
\begin{equation}
m_{\text{fg}}(\bm{\theta}_{\text{fg}}) = \sum_{i=0}^{N_{\text{fg}}-1} a_i \nu^{-2.5+i} = A\bm{\theta}_{\text{fg}},
\label{equation_foreground}
\end{equation}

\noindent where $\nu$ is frequency, $N_{\rm fg}=5$ is the number of foreground terms needed to fit the spectrum over $100$~MHz \citep[][M18]{monsalve2017b}, $A$ is a matrix with columns corresponding to the $\nu^{-2.5+i}$ basis functions, and $\bm{\theta}_{\text{fg}}$ is the vector of foreground polynomial coefficients with elements $a_i$.

We fit Equation~\ref{equation_foreground} to $d-m_{21}(\bm{\theta}_{21})$ using least squares.
The best-fit foreground parameters and model are denoted as $\hat{\bm{\theta}}_{\text{fg}}$ and $\hat{m}_{\text{fg}}$, respectively. The uncertainty of $\hat{\bm{\theta}}_{\text{fg}}$ is encapsulated in their $5\times 5$ covariance matrix, $C=(A^T\Sigma^{-1}A)^{-1}$, where $\Sigma$ is the $N_{\nu}\times N_{\nu}$ covariance matrix of the measured spectrum and $N_{\nu}$ is the number of spectral channels. We construct $\Sigma$ as a diagonal matrix where each element on the diagonal is the sum of the channel variance due to thermal noise and systematic uncertainty. For our channel width of $390.6$~kHz, the standard deviation of the thermal noise is $\approx40$, $6$, and $3$~mK at $90$, $140$, and $190$~MHz, respectively. The systematic uncertainty estimate has a standard deviation of $35$~mK (M18). Finally, the $N_{\nu}\times N_{\nu}$ covariance matrix of $\hat{m}_{\text{fg}}$ is given by $\Sigma_{\text{fg}}=ACA^T$.

With the definitions above, and as derived in M18, the likelihood of the data as a function of $\bm{\theta}_{21}$ after marginalizing over the uncertainty of $\hat{\bm{\theta}}_{\text{fg}}$, is given by 

\begin{align}
&\mathcal{L}(d|\bm{\theta}_{21}) = \sqrt{\frac{(2\pi)^{N_{\text{fg}}-N_{\nu}}}{|\Sigma||C^{-1}|}}\times\\\nonumber
&\exp{\Big\{-\frac{1}{2}\left[d - m_{21}(\bm{\theta}_{21}) - \hat{m}_{\text{fg}}\right]^T(\Sigma + V)^{-1}\left[d - m_{21}(\bm{\theta}_{21}) - \hat{m}_{\text{fg}}\right]\Big\}},
\label{equation_likelihood}
\end{align}

%Here, $N_{\nu}$ is the number of spectral channels, $\Sigma$ is the $N_{\nu}\times N_{\nu}$ covariance matrix of the data, $C=(A^T\Sigma^{-1}A)^{-1}$ is the $5\times 5$ covariance matrix of $\hat{\bm{\theta}}_{\text{fg}}$, $\Sigma_{\text{fg}}=ACA^T$ is the $N_{\nu}\times N_{\nu}$ covariance matrix of $\hat{m}_{\text{fg}}$, and $V = (\Sigma_{\text{fg}}^{-1} - \Sigma^{-1})^{-1}$.

\noindent where $V = (\Sigma_{\text{fg}}^{-1} - \Sigma^{-1})^{-1}$. This likelihood is evaluated for each of the $6.4$ million global signals produced with \texttt{Global21cm}. Defining the prior distribution of the $\bm{\theta}_{21}$ parameters as $\mathcal{P}(\bm{\theta}_{21})$, the 1D and 2D posterior PDFs are obtained by numerically integrating the product $\mathcal{L}(d|\bm{\theta}_{21})\mathcal{P}(\bm{\theta}_{21})$ over the $\bm{\theta}_{21}$ parameters being marginalized. In the EDGES-only analysis we assume a uniform prior distribution for all the parameters over the ranges listed in Table~\ref{table_ranges} (uniform in $\log_{10}$ for $V_{\rm c}$, $f_*$, and $f_{\rm X}$, and in linear scale for the others).

\begin{figure}
\begin{center}
\includegraphics[width=0.48\textwidth]{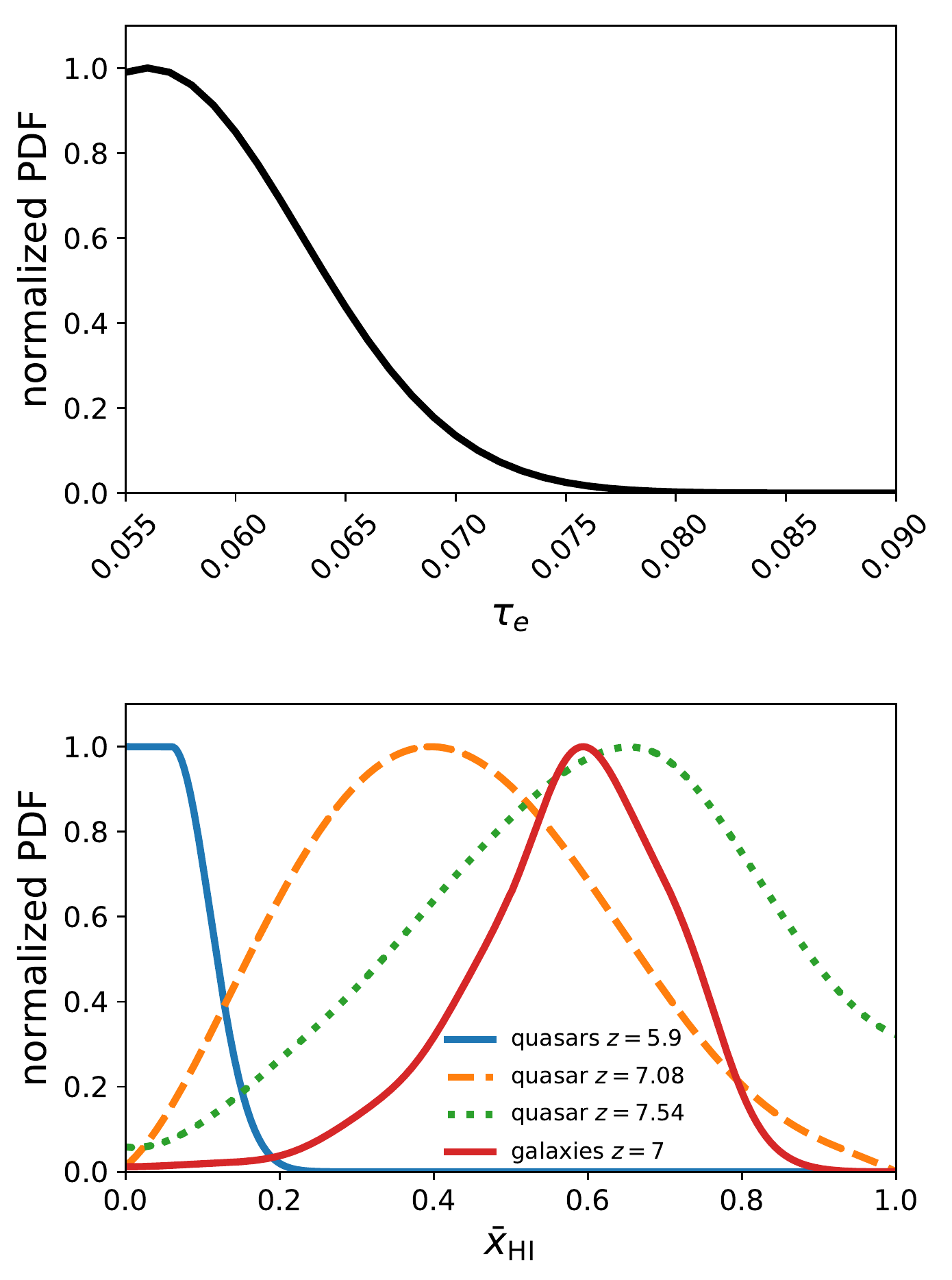}
\caption{PDFs of the external constraints used in our combined analysis. Each PDF is normalized to its peak amplitude. \emph{Top}: PDF of the electron scattering optical depth estimated by \emph{Planck} \citep{planck2018}. We model it as a Gaussian centered at $\tau_{\rm e}=0.056$ and with a width $\sigma=0.007$. \emph{Bottom}: PDFs of the average neutral hydrogen fraction derived from the spectra of high-$z$ quasars and galaxies. The PDF from the quasar at $z=5.9$ was obtained from the fraction of dark Ly-$\alpha$ and Ly-$\beta$ pixels in the quasar spectra \citep{mcgreer2015}. The PDFs from the quasars at $z=7.08$ \citep{greig2017c} and $z=7.54$ \citep{banados2018} were derived from the quasars' Ly-$\alpha$ damping wings. The PDF at $z=7$ \citep{mason2018} was obtained from the analysis of Ly-$\alpha$ transmission from $68$ Lyman Break galaxies.}
\label{figure_external_constraints}
\end{center}
\end{figure}

\subsection{Combined Analysis}
\label{section_analysis_external}

Following \citet{greig2017a} and M18, and improving over M18 by using a more recent prior on $\tau_{\rm e}$ and two additional constraints on $\bar{x}_{\rm H\scriptscriptstyle{I}}$, we derive constraints on the high-$z$ astrophysical parameters after incorporating into our analysis the following external estimates:

\begin{enumerate}
\item $\tau_{\rm e}$ \emph{estimate from Planck}: we use as a prior the estimate $\tau_{\rm e}=0.056 \pm 0.007$ ($68\%$) from \citet{planck2018}, which we model as Gaussian. Although they report several values, here we use this result, which was obtained from the analysis that considers the \emph{Planck} CMB power spectra in combination with CMB lensing reconstruction and baryon acoustic oscillation measurements.

\item $\bar{x}_{\rm H\scriptscriptstyle{I}}$ \emph{constraint at} $z=5.9$: we use the upper limit on $\bar{x}_{\rm H\scriptscriptstyle{I}}$ from \citet{mcgreer2015}, derived from the fraction of pixels that are dark in the Ly-$\alpha$ and Ly-$\beta$ regions of high-$z$ quasar spectra \citep{mesinger2010}. We model this upper limit as a flat probability for $\bar{x}_{\rm H\scriptscriptstyle{I}}\leq0.06$ and a decreasing probability for $\bar{x}_{\rm H\scriptscriptstyle{I}}>0.06$, which follows a Gaussian with center $\bar{x}_{\rm H\scriptscriptstyle{I}}=0.06$ and width $\sigma=0.05$.

\item $\bar{x}_{\rm H\scriptscriptstyle{I}}$ \emph{estimate at} $z=7.08$: we use the $\bar{x}_{\rm H\scriptscriptstyle{I}}$ PDF estimated by \citet{greig2017c} from the Ly-$\alpha$ damping wing analysis of the ULASJ1120+0641 quasar \citep{mortlock2011}. Specifically, we use their result for the `Small H{\scriptsize II}' reionization morphology \citep{mesinger2016}. From this PDF, the $68\%$ estimate is $\bar{x}_{\rm H\scriptscriptstyle{I}}=0.40_{-0.19}^{+0.21}$.

\item $\bar{x}_{\rm H\scriptscriptstyle{I}}$ \emph{estimate at} $z=7.54$: we use the most conservative (i.e., widest) $\bar{x}_{\rm H\scriptscriptstyle{I}}$ PDF estimated by \citet{banados2018} from the Ly-$\alpha$ damping wing analysis of the ULASJ1342+0928 quasar. This estimate accounts for uncertainty in the quasar's intrinsic emission through numerical simulations normalized to the average continuum emission of analog quasars in the Sloan Digital Sky Survey quasar catalog \citep{paris2017}. From this PDF, the $68\%$ estimate is $\bar{x}_{\rm H\scriptscriptstyle{I}}=0.65_{-0.32}^{+0.15}$.

\item $\bar{x}_{\rm H\scriptscriptstyle{I}}$ \emph{estimate at} $z=7$: we use the $\bar{x}_{\rm H\scriptscriptstyle{I}}$ PDF computed by \citet{mason2018} in their analysis of Ly-$\alpha$ transmission from the $68$ LBGs reported by \citet{pentericci2014}. Their analysis incorporates reionization simulations and empirical models of radiative transfer effects in the interstellar medium, yielding the $68\%$ estimate $\bar{x}_{\rm H\scriptscriptstyle{I}} = 0.59^{+0.11 }_{-0.15}$.
\end{enumerate}

% This estimate was obtained from \emph{Planck}-HFI $\ell\leq20$ polarized data in combination with temperature anisotropy data. \textcolor{red}{CHANGE AND USE NEW ESTIMATE 0.056+-0.007 OR IS IT TOO SOON?? PUT SOME OF THIS: 

The PDFs corresponding to these constraints are depicted in Figure \ref{figure_external_constraints}. We consider the \emph{Planck} $\tau_{\rm e}$ PDF a prior because it corresponds to a parameter directly explored in our analysis. This PDF enters into our analysis through $\mathcal{P}(\tau_{\rm e})$. The other constraints are incorporated through an additional likelihood factor, $\mathcal{L}(\bar{x}_{\rm H\scriptscriptstyle{I}}|\bm{\theta}_{21})$, that multiplies the product $\mathcal{L}(d|\bm{\theta}_{21})\mathcal{P}(\bm{\theta}_{21})$. $\mathcal{L}(\bar{x}_{\rm H\scriptscriptstyle{I}}|\bm{\theta}_{21})$ is obtained by evaluating the $\bar{x}_{\rm H\scriptscriptstyle{I}}$ PDFs at the values of $\bar{x}_{\rm H\scriptscriptstyle{I}}$ produced by \texttt{Global21cm} at $z = 5.9$, $7.08$, and $7.54$, for every combination of ${\bm \theta}_{21}$ parameters.

The quasar constraints on $\bar{x}_{\rm H\scriptscriptstyle{I}}$ from \citet{mcgreer2015} and \citet{greig2017c} account for sigthline-to-sightline variance. Specifically, \citet{mcgreer2015} conduct a jackknife analysis where $\bar{x}_{\rm H\scriptscriptstyle{I}}$ is estimated repeatedly after removing one quasar at a time from their $22$-quasar sample. \citet{greig2017c} estimate this effect by computing the $\bar{x}_{\rm H\scriptscriptstyle{I}}$ PDF for $10^5$ sightlines extracted from semi-numerical reionization simulations \citep{mesinger2016}. Recently, \citet{davies2018} conducted an independent analysis of J1120+0641 ($z=7.08$) that accounts for intrinsic emission uncertainty and sightline variance, and obtained the estimate $\bar{x}_{\rm H\scriptscriptstyle{I}}=0.48\pm0.26$ ($68\%$), which is consistent with \citet{greig2017c} at $<1\sigma$. The constraint inferred by \citet{mason2018} from LBGs is tighter than those from J1120+0641 by \citet{greig2017c} and \citet{davies2018} at the same redshift, while consistent at the $\sim1\sigma$ level. This provides support for an ongoing reionization at $z\approx 7$. We choose to include both $z\approx 7$ constraints (J1120+0641 and LBGs) in our analysis because they are completely independent and, when combined, are expected to produce a more  precise and representative estimate of the average fraction of neutral hydrogen.

The constraint from J1342+0928 ($z=7.54$) in \citet{banados2018} used in this paper only accounts for uncertainty in the quasar's intrinsic emission and does not incorporate the effect of sightline variance. However, we still treat this constraint as representative at $z=7.54$ because, in addition to being their most conservative result, the sightline variance for a significantly neutral IGM is expected to be lower than for lower neutral hydrogen fractions \citep{mcgreer2011}. The same quasar was analyzed by \citet{davies2018} including the sightline variance effect. They obtained $\bar{x}_{\rm H\scriptscriptstyle{I}}=0.60_{-0.23}^{+0.20}$ ($68\%$), consistent with \citet{banados2018}.

During the preparation of this manuscript, \citet{greig2019} presented a new analysis of J1342+0928, which also accounts for the sightline variance effect. Unlike \citet{banados2018} and \citet{davies2018}, they do not find evidence for a significantly neutral IGM at $z=7.54$. Their best estimates for $\bar{x}_{\rm H\scriptscriptstyle{I}}$ --- which depend on the reionization morphology assumed --- are below $0.3$ and consistent with zero at $\lesssim 1\sigma$. However, since the $\bar{x}_{\rm H\scriptscriptstyle{I}}$ PDFs are wide, these results are in tension with \citet{banados2018} and \citet{davies2018} only at $\lesssim 1.5\sigma$. Here we use the conservative estimate from \citet{banados2018} and leave for future work incorporating newer quasar constraints on $\bar{x}_{\rm H\scriptscriptstyle{I}}$.

\section{Results}
\label{section_results}

\begin{figure*}
\begin{center}
\includegraphics[width=0.995\textwidth]{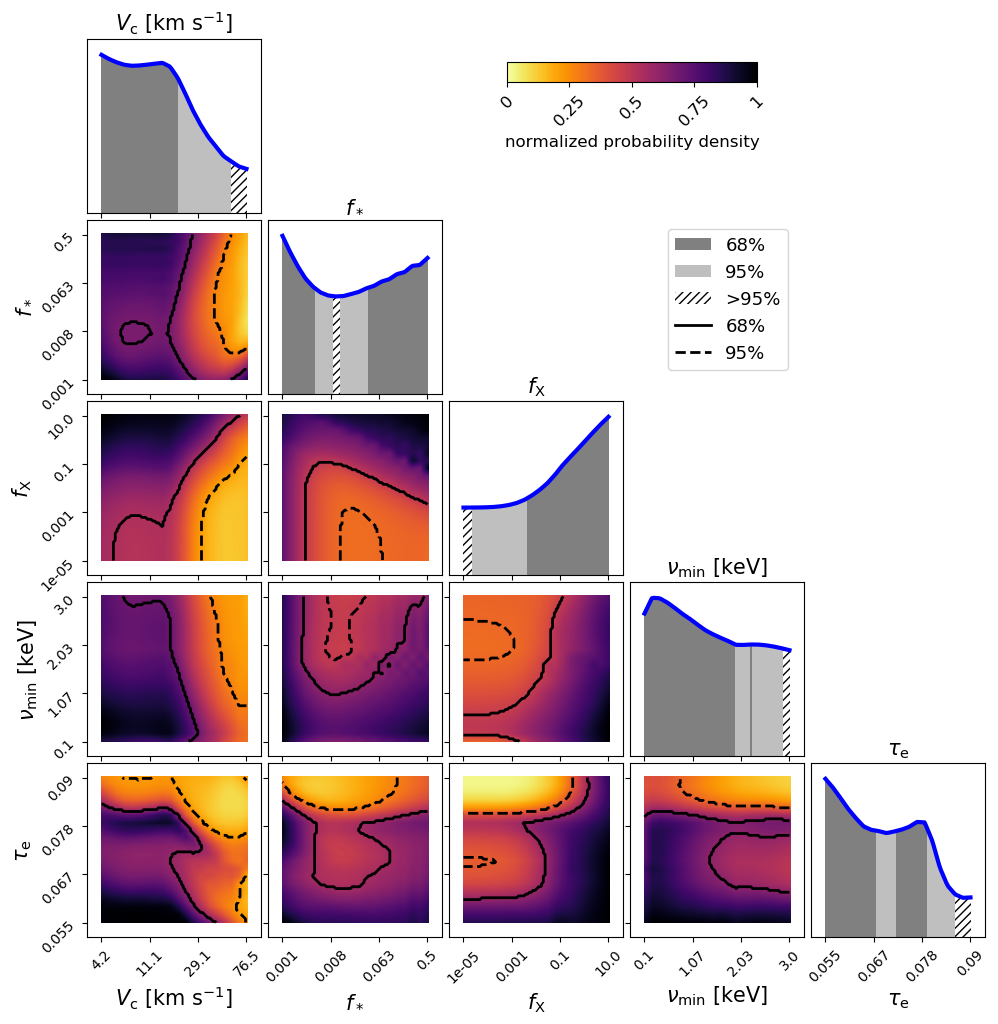}
\caption{PDFs of the astrophysical parameters derived from the analysis of the EDGES High-Band spectrum alone \citep{monsalve2017b} assuming fixed $R_{\text{mfp}}=30$~Mpc. The regions of parameter space that are disfavored by EDGES (depicted as hatched and light gray bands in the 1D PDFs and as yellow areas on the 2D PDFs) are those of high $V_{\rm c}$, intermediate $f_{\star}$, low $f_\text{X}$, high $\nu_{\rm min}$, and high $\tau_{\rm e}$. The marginalized $68\%$ and $95\%$ limits obtained from this analysis are listed in Table~\ref{table_limits} as case A, as well as in Table~\ref{table_taue} for $\tau_{\rm e}$.}
\label{figure_vmin}
\end{center}
\end{figure*}

\begin{figure}
\begin{center}
\includegraphics[width=0.45\textwidth]{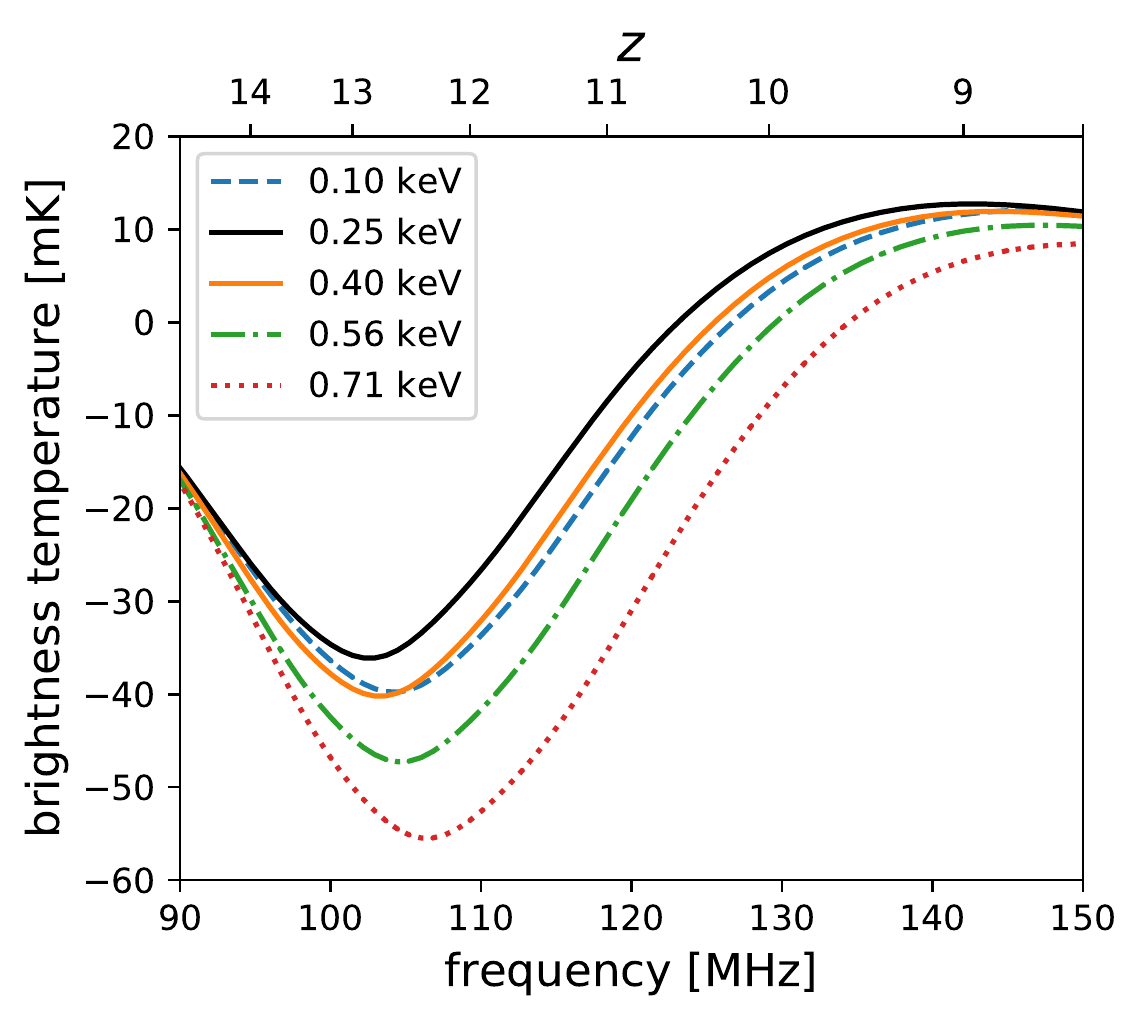}
\caption{Variation of the global signal for low values of $\nu_{\rm min}$. Over most of the parameter range explored in this paper, as $\nu_{\rm min}$ decreases, the absorption amplitude also decreases and the absorption peak is shifted to lower frequencies. However, this trend is reversed below $\nu_{\text{min}}\approx0.25$ keV as seen in this figure. The signals shown here as examples correspond to $V_{\rm c}=76.5$~km s$^{-1}$, $f_*=0.026$, $f_{\text{X}}=10$, $\tau_{\rm e}=0.06$, and $R_{\text{mfp}}=31$~Mpc.}
\label{figure_signals_vmin}
\end{center}
\end{figure}

Now we present the constraints on the six astrophysical parameters derived from the analysis of (1) the EDGES data, (2) the external constraints from \emph{Planck}+quasars+galaxies, and (3) the combination EDGES+\emph{Planck}+quasars+galaxies. In particular, we show results for (2) because the external constraints have a significant impact on $V_{\rm c}$ and $R_{\rm mfp}$, in addition to $\tau_{\rm e}$, and we want to highlight these results independently. The results for the case where $\nu_{\rm min}$ is the fifth parameter in the analysis are shown in Figures \ref{figure_vmin}, \ref{figure_vmin_external}, and \ref{figure_vmin_combined}. The results for $R_{\rm mfp}$ when treated as the fifth parameter are shown in Figure \ref{figure_rmfp}. In Figure \ref{figure_rmfp} we do not show the PDFs that only involve the other four parameters, as they are similar to those in Figures \ref{figure_vmin}, \ref{figure_vmin_external} and \ref{figure_vmin_combined}. Table~\ref{table_limits} presents the marginalized $68\%$ and $95\%$ limits on all the parameters from the EDGES-only and combined analyses. In Table~\ref{table_taue} we show the estimates for $\tau_{\rm e}$ derived from each individual observation, as well as for different combinations. Unless stated otherwise, the limits quoted for reference in the rest of this Section correspond to the case with $\nu_{\rm min}$ as fifth parameter.

%  The main effect of the external constraints compared to Figure~\ref{figure_vmin} is a significant reduction of the high-probability range in the joint PDF of $V_{\rm c}-\tau_{\rm e}$. 
% This results in the marginalized estimate $\tau_{\rm e}=0.062^{+0.005}_{-0.003}$ ($68\%$). The constraints on the other parameters remain mostly determined by the EDGES High-Band spectrum. 

% High-amplitude global signals that vary strongly within the band are incompatible with the EDGES High-Band spectrum \citep{monsalve2017b}. 

\subsection{EDGES-only Analysis}
As we can see in the 1D and 2D PDFs of Figure \ref{figure_vmin} and the top row of Figure \ref{figure_rmfp}, the EDGES High-Band measurement provides significant discrimination across the explored parameter space. \citet{monsalve2017b} showed that the High-Band data are incompatible with global signals that have high amplitude and vary rapidly within the band. For our models and parametrization, this translates into the disfavoring of models with high $V_{\rm c}$, intermediate $f_*$, low $f_{\rm X}$, and high $\tau_{\rm e}$. Models with high $\nu_{\rm min}$ and high $R_{\rm mfp}$ are also disfavored, although the data are less sensitive to variations in these parameters due to their weaker impact on the global signal. From the PDFs, we derive the following constraints on each one of the parameters:
% This enables us to determine  values of astrophysical parameters
% The constraints on each parameter are as follows:

\begin{itemize}[leftmargin=*, label={}]

\item $V_{\rm c}$: Along with $f_*$, $V_{\rm c}$ determines the timing of the Ly-$\alpha$ coupling and drives the evolution of the signal all the way to the onset of heating, affecting the location and depth of the absorption trough.  As \citet{cohen2017, cohen2019} indicate (e.g., Fig. 6 of the latter paper),  Ly-$\alpha$ coupling is predicted to take place at $z\gtrsim 20$ and, thus, cannot be directly probed by EDGES High-Band. However, the High-Band data are sensitive to the features of the absorption trough and, as a result, can discriminate between low and high values of $V_{\rm c}$. As we see in the PDFs, high values of $V_{\rm c}$ (i.e., higher minimum mass of star-forming haloes) are disfavored because they result in narrower troughs centered at higher frequencies, to which EDGES has higher sensitivity.  We rule out $V_{\rm c} > 19.3$~km~s$^{-1}$ at $68\%$ confidence. This velocity threshold is close to the limit of atomic cooling and corresponds to the minimum halo mass of $1.3\times 10^8$~M$_{\odot}$ at $z=10$. At $95\%$ confidence we rule out $V_{\rm c} > 56$~km~s$^{-1}$, which corresponds to $M_{\rm min} \sim 3.1\times 10^9$~M$_{\odot}$ at $z= 10$.

% The PDFs of $V_{\rm c}$ show that EDGES disfavors high values of this parameter.

\item $f_*$: Low values of $f_*$ result in inefficient Ly-$\alpha$ coupling and, as a result,  shallow absorption profiles, while high values produce deeper but wider absorption profiles. The EDGES High-Band data and modeling provide low sensitivity to both types of signals, which results in a high probability assigned to low and high $f_*$. On the other hand, our analysis disfavors intermediate values of $f_*$, which produce sharper signatures in the High-Band range. Specifically, we rule out $0.4\% < f_* <3.9\%$ ($68\%$).

\item $f_{\rm X}$: The X-ray heating efficiency is one of the parameters that control the location of the absorption minimum and the high-frequency slope of the trough. A higher $f_{\rm X}$ results in sharper but shallower troughs centered at lower frequencies owing to more efficient heating, and could also result in a significant emission feature during reionization. A lower $f_{\rm X}$ produces deeper and wider troughs centered at higher frequencies, as well as a suppressed or vanishing emission signal. The EDGES spectrum is more sensitive to low-$f_{\rm X}$ signals, although the high sensitivity expected from their large depth is compensated by the lower sensitivity due to the larger width. As we can see in the PDFs, low values of $f_{\rm X}$ are disfavored for most of the parameter combinations. After marginalization, we rule out $f_{\rm X}<0.0042$ ($2 \times 10^{-5}$) at $68\%$ ($95\%$) confidence.

\item $\nu_{\rm min}$: Although the global signal is less sensitive to changes in $\nu_{\rm min}$ compared to the previous parameters, EDGES High-Band can still discriminate across the range explored. Specifically, EDGES disfavors high values of $\nu_{\rm min}$, corresponding to harder X-ray SEDs that produce wider and deeper absorption troughs shifted to higher frequencies. As $\nu_{\rm min}$ decreases, the fraction of soft X-rays emitted by sources increases, which results in more efficient IGM heating and in earlier and shallower absorption troughs. Our conservative $68\%$ upper limit is $\nu_{\rm min}=2.3$~keV; however, as can be seen in the 1D $\nu_{\rm min}$ PDF and in Table~\ref{table_limits}, this limit accounts for the narrow range $2.2-2.3$~keV that also falls within the $68\%$ limits. Ignoring this range we obtain the limit $\nu_{\rm min}=1.9$~keV ($68\%$). In the 1D PDF we also notice that the probability has a peak at $\nu_{\text{min}}\approx 0.25$~keV and that it decreases for lower values. We explore the origin of this feature in Figure~\ref{figure_signals_vmin}; we see that, for $\nu_{\rm min}>0.25$~keV, and as $\nu_{\rm min}$ decreases, the absorption trough in the $21$~cm signal becomes shallower and the center is shifted to lower frequencies. However, as values reach and decrease below $\nu_{\text{min}}\approx0.25$~keV, the absorption becomes deeper again and the center is shifted to higher frequencies, approaching the shapes observed for $\nu_{\rm min}>0.25$~keV. This reversed dependence of the global signal below a $\nu_{\rm min}$ threshold is due to an effective hardening of the X-ray SED at low $\nu_{\rm min}$, as most of the energy produced by the sources is deposited very close to the star-forming regions \citep[see Section 2.2.5 of][]{greig2017b}. The 1D $\nu_{\rm min}$ PDF reflects that EDGES High-Band has the lowest constraining capability around this threshold.

\item $\tau_{\rm e}$: Changes in $\tau_{\rm e}$ affect the evolution of the IGM ionized hydrogen fraction. Higher values imply higher ionizing efficiency of sources, which leads to an earlier reionization and a global signal with a shallower but narrower absorption feature and a weaker emission peak. Lower values of $\tau_{\rm e}$ result in delayed, deeper, and wider troughs, as well as in a potentially stronger emission feature that peaks at higher frequencies. The general trend in the $\tau_{\rm e}$ PDFs from EDGES is a probability density that decreases for higher $\tau_{\rm e}$, which is consistent with the preference of low $\tau_{\rm e}$ by \emph{Planck}. Our conservative $68\%$ upper limit, accounting for the high-probability bump centered at $\tau_{\rm e}\approx 0.076$ (described in the next paragraph), is $\tau_{\rm e}=0.080$. We note, however, that due to the higher noise at the low-end of the spectrum ($\lesssim 110$~MHz), the sensitivity of the EDGES High-Band data to models with $\tau_{\rm e}\gtrsim 0.09$ decreases significantly. Higher-sensitivity measurements at $\lesssim 110$~MHz, such as those provided by EDGES Low-Band, are required to access these higher optical depths.

% in the context of the absorption feature reported by \citet{bowman2018}, 

\begin{figure*}
\begin{center}
\includegraphics[width=0.995\textwidth]{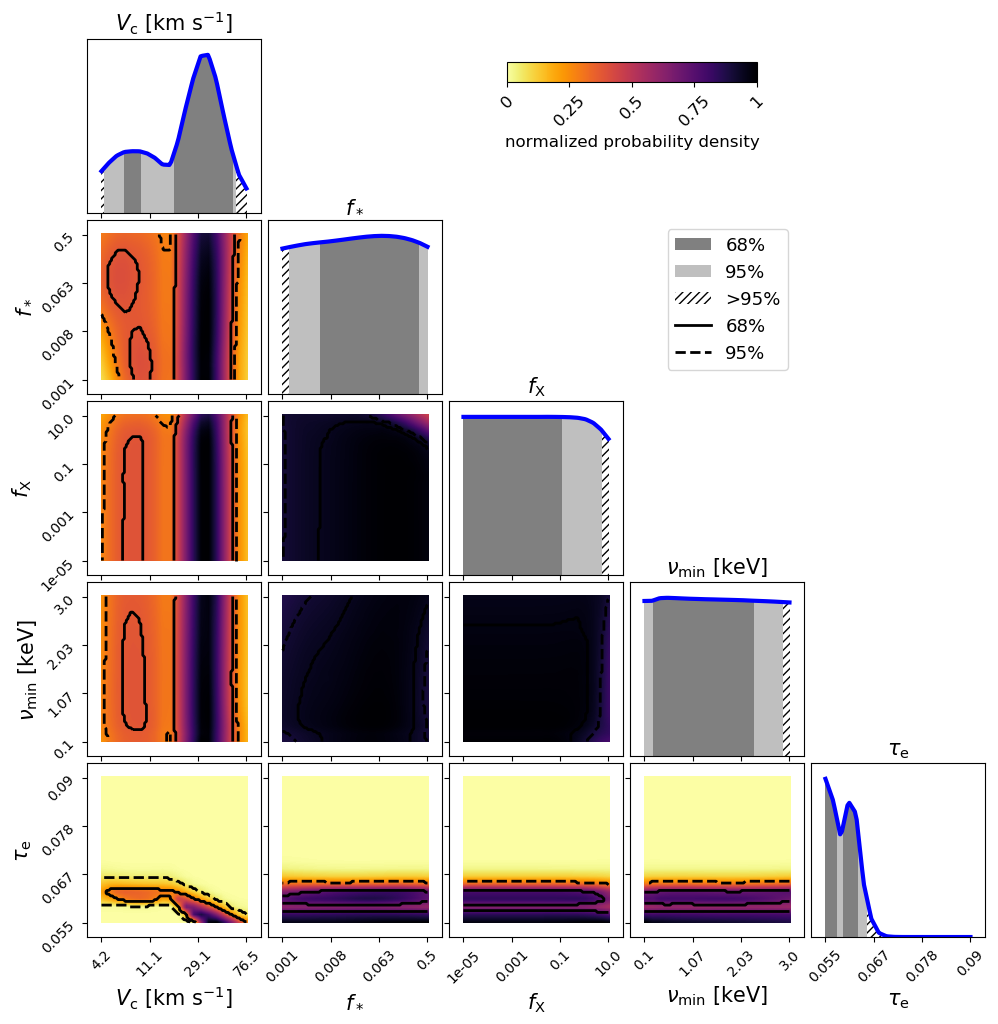}
\caption{PDFs of the astrophysical parameters derived from the analysis of the external constraints alone and assuming fixed $R_{\text{mfp}}=30$~Mpc. The external constraints correspond to a prior on $\tau_{\rm e}$ from \emph{Planck} \citep{planck2018} and estimates for $\bar{x}_{\rm H\scriptscriptstyle{I}}$ from high-$z$ quasars \citep{mcgreer2015, greig2017c, banados2018} and Lyman Break galaxies \citep{mason2018}. The external constraints are described in Section~\ref{section_analysis_external} and depicted in Figure~\ref{figure_external_constraints}. An important result of this analysis is the strong joint constraint on $\tau_{\rm e}-V_{\rm c}$. The marginalized results for $\tau_{\rm e}$ are summarized in Table~\ref{table_taue}.}
\label{figure_vmin_external}
\end{center}
\end{figure*}

Beyond the main trend, in Figures \ref{figure_vmin} and \ref{figure_rmfp} (top row) we see that the $\tau_{\rm e}$ PDFs have the most irregular structure among the parameters. When projected onto the 1D $\tau_{\rm e}$ PDF, this structure is seen as a bump at $\tau_{\rm e}\approx0.076$. To understand its origin we compute the PDFs for simulated EDGES spectra. These spectra are produced starting from the five-term foreground model that best fits the measured spectrum, to which we add noise drawn from the same noise profile as the measurement. We also add ripples that mimic those observed in the measured spectrum above the foreground model \citep[see Figure 4 of][]{monsalve2017b}. In some cases we add ripples only within sub-bands of the spectrum in order to evaluate their specific effect. We find that the bump at $\tau_{\rm e}\approx0.076$ is produced by $21$~cm signals that match ripples in the measured spectrum within the range $\approx125-145$~MHz. Simulations without these ripples produce PDFs that decrease smoothly with $\tau_{\rm e}$, without a bump at $\approx0.076$. Future re-processing and modeling of the High-Band data might reveal the origin of the ripples. New measurements with different instruments could be used to revise the PDFs of this parameter. As seen in the 1D PDF of Figure~\ref{figure_vmin}, the bump at $\tau_{\rm e}\approx0.076$ represents a second range contained within the $68\%$ confidence limits, in addition to the larger range at low $\tau_{\rm e}$. Ignoring the bump and considering only the low $\tau_{\rm e}$ region, the $68\%$ limit is $\tau_{\rm e}=0.067$.

\item $R_{\rm mfp}$: The sensitivity of the global signal to changes in $R_{\rm mfp}$ is lower than for the other parameters and comparable to that for $\nu_{\rm min}$. Higher values of $R_{\rm mfp}$ correspond to a faster growth of ionized bubbles, and, thus, to a faster reionization process and a sharper end of reionization \citep{greig2017a, greig2017b}. The 1D $R_{\rm mfp}$ PDF in the top row of Figure \ref{figure_rmfp} shows that the EDGES spectrum disfavors higher values of $R_{\rm mfp}$. We rule out $R_{\text{mfp}}>36$~Mpc ($68\%$) for fixed $\nu_{\text{min}}=0.5$~keV.

\end{itemize}

In Table~\ref{table_limits}, the constraints from the EDGES-only analyses are presented as cases A (for fixed $R_{\rm mfp}=30$~Mpc) and B (for fixed $\nu_{\rm min}=0.5$~keV).

\subsection{External Constraints}
\label{section_results_external}

Here we describe the constraints on the astrophysical parameters derived from the $\tau_{\rm e}$ estimate from \emph{Planck}, the $\bar{x}_{\rm H\scriptscriptstyle{I}}$ estimates from quasars at $z=5.9$, $7.08$, and $7.54$, and the $\bar{x}_{\rm H\scriptscriptstyle{I}}$ estimate from galaxies at $z=7$. These external estimates characterize the evolution of the neutral hydrogen fraction and, therefore, strongly constrain the reionization parameters, i.e., $\tau_{\rm e}$ and  $R_{\rm mfp}$. However, due to the correlation between reionization, the star formation history, and --- to a lesser degree --- heating,  there is some degeneracy between the reionization parameters and $V_{\rm c}$, as well as a much weaker degeneracy with $f_*$, $f_{\rm X}$, and $\nu_{\rm min}$. As we show in this section, these degeneracies are reflected in the parameters constraints. The results of the analyses that combine all the external estimates are presented in Figure~\ref{figure_vmin_external} and the middle row of Figure~\ref{figure_rmfp}.

Consider first the limits from the external estimates on $\tau_{\rm e}$, summarized in Table~\ref{table_taue}. The upper limit $\bar{x}_{\rm H\scriptscriptstyle{I}}\leq 0.06+0.05$~($68\%$) at $z=5.9$ from \citet{mcgreer2015} significantly reduces the probability of late reionization and, hence, of low $\tau_{\rm e}$. From this constraint alone we derive the marginalized lower limit $\tau_{\rm e}>0.068$ ($0.056$) at $68\%$ ($99\%$) confidence. On the other hand, quasars at $z\gtrsim 7$ suggest that the IGM was significantly neutral at these redshifts, with  $\bar{x}_{\rm H\scriptscriptstyle{I}}=0.40_{-0.19}^{+0.21}$ \citep[$68\%$,][]{greig2017c} and  $\bar{x}_{\rm H\scriptscriptstyle{I}}=0.65_{-0.32}^{+0.15}$ \citep[$68\%$,][]{banados2018} measured at $z = 7.08$ and $z=7.54$, respectively. These data complement the upper limit from \citet{mcgreer2015} by disfavoring an early reionization and, thus, high values of $\tau_{\rm e}$. Specifically, both measurements independently impose the upper limit $\tau_{\rm e}<0.065$ ($68\%$). Finally, the tighter constraint $\bar{x}_{\rm H\scriptscriptstyle{I}} = 0.59^{+0.11 }_{-0.15}$~($68\%$) from LBGs at $z=7$ gives preference to lower optical depths than the quasars. Analyzing this constraint alone results in the upper limit $\tau_{\rm e}<0.061$ ($68\%$), which we notice is in mild tension with the lower limit from \citet{mcgreer2015}.

Joint analysis of the three quasar constraints favors $\tau_{\rm e}$ in the range $0.057-0.067$ ($68\%$). This range is tighter than current estimates from \emph{Planck}, and is also in mild tension with those ($\lesssim 2\sigma$, depending on the specific estimate) since the \emph{Planck} best fits lie below our quasar $68\%$ range \citep{planck2018}. Combining the quasar and LBG neutral fraction constraints we obtain the upper limit $\tau_{\rm e}<0.064$~($68\%$). Here, the tight neutral fraction constraint from LBGs has increased the consistency between this combined $\tau_{\rm e}$ result and the $\tau_{\rm e}$ estimates from \emph{Planck}. This is noteworthy considering that the assumed ionization histories are different; \emph{Planck} uses a `tanh' phenomenological dependence of $\bar{x}_{\rm H\scriptscriptstyle{I}}$ on redshift, while we used the realistic neutral fractions produced by \texttt{Global21cm} to derive the quasar and LBG constraints. Finally, incorporating the \emph{Planck} prior ($\tau_{\rm e}=0.056 \pm 0.007$) we obtain $\tau_{\rm e}<0.063$~($68\%$). This latter result, derived from the combination of our five external constraints, is the one corresponding to the 1D PDF of Figure~\ref{figure_vmin_external}. In this PDF we see a probability dip in the middle of an otherwise smooth trend of increasing probability toward low values. The dip has the effect of excluding the narrow range $\tau_{\rm e}=0.058-0.059$ from the $68\%$ probability region. This feature is explained by the combination of two factors: (1) as we pointed out above, there is a mild tension between the lower values of $\tau_{\rm e}$ favored by LBGs (and \emph{Planck}), and the higher $\tau_{\rm e}$ preferred by the quasars; (2) the piecewise dependence of the SFE on $V_{\rm c}$, as implemented in \texttt{Global21cm}, which induces features in the PDFs (more details below). Considering that the $\tau_{\rm e}$ values preferred by the combined constraints are low and reach our current low-end optical depth cutoff, even when including the upper limit on $\bar{x}_{\rm H\scriptscriptstyle{I}}$ from \citet{mcgreer2015}, we plan to extend the parameter range to values below $\tau_{\rm e}=0.055$ in future versions of \texttt{Global21cm}.

%The marginalized PDF suggests a high probability below the low-end optical depth cutoff of $\tau_{\rm e}=0.055$ implemented in \texttt{Global21cm}. It also shows a probability dip that excludes the narrow range $\tau_{\rm e}=0.057-0.059$ from the $68\%$ region.

The joint constraints on $\tau_{\rm e}$ and $V_{\rm c}$ obtained when we apply all the external estimates are shown in the corresponding 2D PDF of Figure~\ref{figure_vmin_external}. This PDF reflects the degeneracy between these two parameters in their effect on the global reionization history. For a fixed $\tau_{\rm e}$, reionization is slower in the case of low $V_{\rm c}$. In particular, for low $V_{\rm c}$ the tail of $\bar{x}_{\rm H\scriptscriptstyle{I}}$ at the end of reionization is longer and, therefore, the values of $\bar{x}_{\rm H\scriptscriptstyle{I}}$ at a fixed redshift are higher than in the case of higher $V_{\rm c}$. Hence, for low $\tau_{\rm e}$ the scenarios with lower $V_{\rm c}$ are more likely to violate the upper limit on $\bar{x}_{\rm H\scriptscriptstyle{I}}$ at $z = 5.9$. On the other hand, to keep a sufficiently high $\bar{x}_{\rm H\scriptscriptstyle{I}}$ as required at $z\gtrsim 7$, the  constraints on the neutral fraction prefer low $V_{\rm c}$ at low $\tau_{\rm e}$. As a result, the high-probability region in the $\tau_{\rm e} - V_{\rm c}$ 2D PDF is confined to a narrow band that is mainly produced by the complementary effects of the $\bar{x}_{\rm H\scriptscriptstyle{I}}$ constraint at $z = 5.9$ and those at $z\gtrsim 7$. At $V_{\rm c}<16.5$~km~s$^{-1}$, the band is centered at $\tau_{\rm e}\approx0.064$ and only has a weak dependence on $V_{\rm c}$. At $V_{\rm c}=16.5$~km~s$^{-1}$ the band goes through a knee and, for higher $V_{\rm c}$, $\tau_{\rm e}$ decreases for increasing $V_{\rm c}$. Since the \emph{Planck} prior and the $\bar{x}_{\rm H\scriptscriptstyle{I}}$ estimate from LBGs prefer lower $\tau_{\rm e}$, the highest probability along the band occurs for high $V_{\rm c}$. The sharpness of the knee is not physical; it is an artifact of our models produced by the piecewise SFE of Equation \ref{equation_fstar}, which changes the trend exactly at $V_{\rm c}=16.5$~km~s$^{-1}$, corresponding to the atomic cooling threshold. We observe that, after marginalization, the knee results in relatively sharp features in the 1D PDFs of $V_{\rm c}$ and $\tau_{\rm e}$; specifically, probability dips at $V_{\rm c}\approx 16.5$~km~s$^{-1}$ and $\tau_{\rm e}\approx 0.058$. We plan to improve the $V_{\rm c}$ transition in future modeling.

In the 1D $V_{\rm c}$ PDF it is more evident that, unlike the EDGES data, the external constraints prefer high values of $V_{\rm c}$. This PDF is dominated by a bump that contains most of the $68\%$ probability volume and peaks at $V_{\rm c}\approx 35$~km~s$^{-1}$. The $68\%$ lower limit of the bump is $V_{\rm c}=17.9$~km~s$^{-1}$, i.e., close to the atomic cooling threshold, and the upper limit is $58.1$~km~s$^{-1}$, which corresponds to $M_{\rm min} \sim 3.5\times 10^9$~M$_{\odot}$ at $z= 10$.

At fixed $\tau_{\rm e}$ and $V_{\rm c}$, the 2D PDFs of $f_*$, $f_{\rm X}$, and $\nu_{\rm min}$ are nearly flat, reflecting the small effect of these parameters on $\bar{x}_{\rm H\scriptscriptstyle{I}}$. The contribution of X-rays to reionization is non-negligible, however. This can be appreciated better after marginalization, as a mild preference of the data for high $f_*$, low $f_{\rm X}$, and low $\nu_{\rm min}$, in their 1D PDFs.

As the middle row of Figure~\ref{figure_rmfp} shows, the external constraints favor high values of $R_{\rm mfp}$, which (like high $V_{\rm c}$) correspond to a faster reionization. This is opposite to the preference by EDGES data, and results from the need to simultaneously satisfy the neutral fraction upper limit at $z=5.9$ and the high neutral fraction at $z\gtrsim 7$, as well as produce a low optical depth.

\begin{figure*}
\begin{center}
\includegraphics[width=0.995\textwidth]{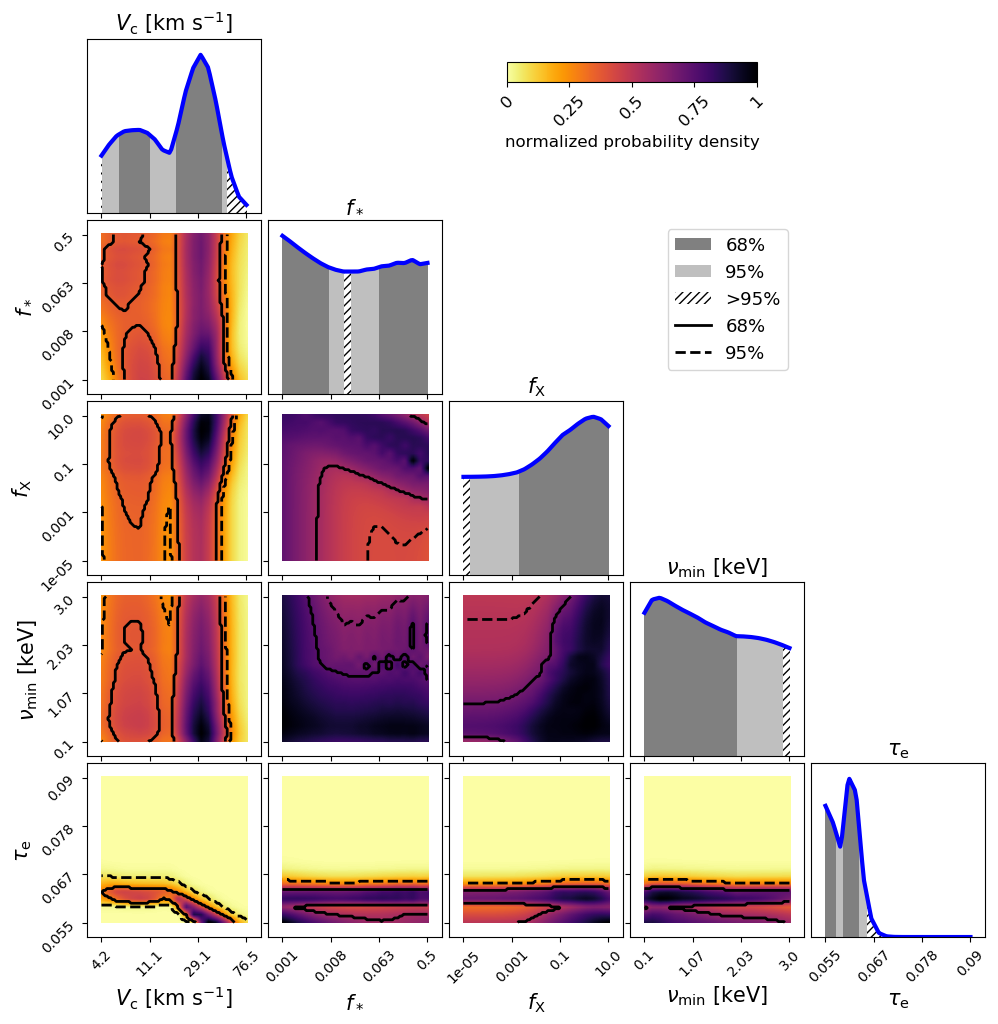}
\caption{PDFs of the astrophysical parameters derived from the combined analysis that includes the EDGES High-Band spectrum \citep{monsalve2017b} and external estimates for $\tau_{\rm e}$ and $\bar{x}_{\rm H\scriptscriptstyle{I}}$ \citep{mcgreer2015, greig2017c, banados2018, mason2018, planck2018}. Here we assume a fixed $R_{\text{mfp}}=30$~Mpc. Comparing the 1D PDFs in this figure with those in Figures~\ref{figure_vmin} and \ref{figure_vmin_external}, we see that (1) EDGES drives the constraints on $f_*$, $f_{\rm X}$, and $\nu_{\rm min}$, (2) the external observations drive the constraint on $\tau_{\rm e}$, and (3) EDGES and the external observations impact significantly the constraint on $V_{\rm c}$. The marginalized $68\%$ and $95\%$ limits from this analysis are listed in Table~\ref{table_limits} as case C, as well as in Table~\ref{table_taue} for $\tau_{\rm e}$.}
\label{figure_vmin_combined}
\end{center}
\end{figure*}

\begin{figure*}
\begin{center}
\includegraphics[width=0.995\textwidth]{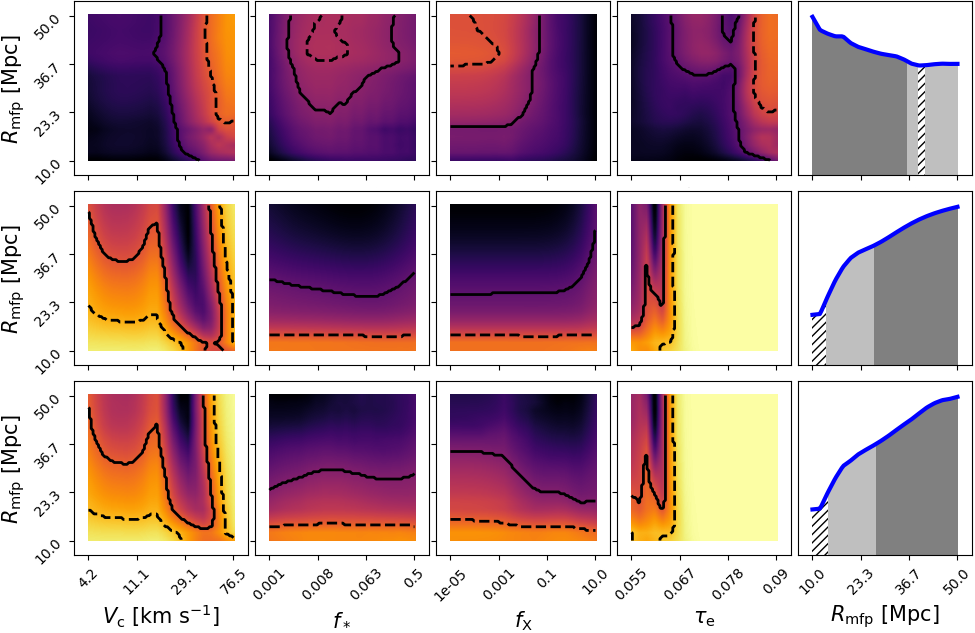}
\caption{PDFs of $R_{\text{mfp}}$ derived from the analysis of (\emph{top row}) the EDGES High-Band spectrum \citep{monsalve2017b}, (\emph{middle row}) the external estimates for $\tau_{e}$ and $\bar{x}_{\rm H\scriptscriptstyle{I}}$, from \emph{Planck} and high-$z$ quasars and galaxies \citep{mcgreer2015, greig2017c, banados2018, mason2018, planck2018}, and (\emph{bottom row}) the combination EDGES + \emph{Planck} + quasars + galaxies. For these results we have fixed $\nu_{\text{min}}=0.5$~keV. EDGES data alone disfavor high values of $R_{\text{mfp}}$ (top row, rightmost column), which correspond to global signals with sharper features. The external constraints favor high values of $R_{\text{mfp}}$ (middle row, rightmost column). When all the observations are combined, the external constraints have the strongest influence and high values of $R_{\text{mfp}}$ remain preferred (bottom row, rightmost column).}
\label{figure_rmfp}
\end{center}
\end{figure*}

\capstartfalse
\begin{deluxetable}{ccrrrr}
\tabletypesize{\scriptsize}
\tablewidth{0pt}
\tablecaption{Marginalized $68\%$ and $95\%$ parameter limits \label{table_limits}   }
\tablehead{& & \multicolumn{2}{c}{$68\%$} & \multicolumn{2}{c}{$95\%$} \\     
\cmidrule(lr){3-4}  \cmidrule(lr){5-6} \colhead{Parameter} & \colhead{Case} & \colhead{Min} & \colhead{Max} & \colhead{Min} & \colhead{Max}}
\startdata
\\
$V_{\rm c}$ [km\;s$^{-1}$]                                 & \emph{A} \dotfill & $4.2$  & $19.3$ & $4.2$ & $56.0$ \\
\\
                                                           & \emph{B} \dotfill & $4.2$  & $21.5$ & $4.2$ & $58.1$ \\
\\
                                                           & \emph{C} \dotfill & $6.0$  & $11.1$ & $4.2$ & $52.0$ \\
                                                           &   & $18.6$ & $46.6$ &       &        \\
\\
                                                           & \emph{D} \dotfill & $6.0$  & $11.1$ & $4.2$ & $52.0$ \\
                                                           &   & $17.9$ & $46.6$ &       &        \\
\\
\hline
\\
$f_*$                                  & \emph{A} \dotfill & $0.001$ & $0.004$ & $0.001$ & $0.009$ \\
                                       &   & $0.039$ & $0.5$   & $0.012$ & $0.5$   \\
\\
                                       & \emph{B} \dotfill & $0.001$ & $0.004$ & $0.001$ & $0.011$ \\
                                       &   & $0.036$ & $0.5$   & $0.015$ & $0.5$   \\
\\
                                       & \emph{C} \dotfill & $0.001$ & $0.007$ & $0.001$ & $0.014$ \\
                                       &   & $0.062$ & $0.5$   & $0.019$ & $0.5$   \\
\\
                                       & \emph{D} \dotfill & $0.001$ & $0.009$ & $0.001$ & $0.38$ \\
                                       &   & $0.036$ & $0.045$ &         &    \\
                                       &   & $0.053$ & $0.324$ &         &    \\
\\
\hline
\\
$f_{\rm X}$                            & \emph{A} \dotfill & $0.0042$ & $10$ & $2\times 10^{-5}$ & $10$ \\
\\
                                       & \emph{B} \dotfill & $0.0025$ & $10$ & $2\times 10^{-5}$ & $10$ \\
\\
                                       & \emph{C} \dotfill & $0.0021$ & $10$ & $2\times 10^{-5}$ & $10$ \\
\\
                                       & \emph{D} \dotfill & $0.0012$ & $10$ & $2\times 10^{-5}$ & $10$ \\
\\
\hline
\\
$\nu_{\rm min}$\;[keV]                 & \emph{A} \dotfill & $0.1$ & $1.9$ & $0.1$ & $2.9$ \\
                                       &   & $2.2$ & $2.3$ &       &        \\
\\
                                       & \emph{C} \dotfill & $0.1$ & $2.0$ & $0.1$ & $2.9$ \\
\\
\hline
\\
$\tau_{\rm e}$                         & \emph{A} \dotfill & $0.055$ & $0.067$ & $0.055$ & $0.086$ \\
                                       &   & $0.072$ & $0.080$ &         &         \\
\\
                                       & \emph{B} \dotfill & $0.055$ & $0.072$ & $0.055$ & $0.087$ \\
                                       &   & $0.074$ & $0.079$ &         &         \\
\\
                                       & \emph{C} \dotfill & $0.055$ & $0.057$ & $0.055$ & $0.065$ \\
                                       &   & $0.059$ & $0.063$ &         &         \\
\\
                                       & \emph{D} \dotfill & $0.055$ & $0.057$ & $0.055$ & $0.065$ \\
                                       &   & $0.059$ & $0.063$ &         &         \\
\\
\hline
\\
$R_{\rm mfp}$\;[Mpc]                   & \emph{B} \dotfill & $10.0$ & $36.1$ & $10.0$ & $39.1$ \\
                                       &   &      &      & $41.1$ & $50.0$ \\
\\
                                       & \emph{D} \dotfill & $27.5$ & $50.0$ & $14.3$ & $50.0$ \\
\enddata
\tablecomments{\\1) Cases: (A) EDGES only, $R_{\rm mfp}=30$~Mpc; (B) EDGES only, $\nu_{\text{min}}=0.5$~keV; (C) Combined constraints, $R_{\rm mfp}=30$~Mpc; (D) Combined constraints, $\nu_{\text{min}}=0.5$~keV.\\2) For some parameters, a given probability volume ($68\%$ or $95\%$) is contained within two or three disjoint value ranges. These ranges are presented in the table as rows associated with the same `Case' letter.}
\end{deluxetable}

\subsection{Combined Analysis}
\label{section_results_combined}

Next, we present the astrophysical constraints obtained in the analysis that includes the EDGES High-Band data and the external estimates for the optical depth and the neutral hydrogen fraction. Compared to the results from EDGES alone, the external estimates have the strongest impact on the PDFs of $\tau_{\rm e}$, $V_{\rm c}$, and $R_{\rm mfp}$, while the PDFs of $f_*$, $f_{\rm X}$, and $\nu_{\rm min}$ are mainly determined by EDGES. The results are shown in Figure~\ref{figure_vmin_combined} (for fixed $R_{\rm mfp}=30$~Mpc) and in the bottom row of Figure~\ref{figure_rmfp} (for fixed $\nu_{\text{min}}=0.5$~keV). They are also summarized in Table~\ref{table_limits} (as cases C and D) and in Table~\ref{table_taue} for $\tau_{\rm e}$.

Considering the 2D PDF for $\tau_{\rm e}$ and $V_{\rm c}$ in Figure~\ref{figure_vmin_combined}, we see that the narrow high-probability band at low $\tau_{\rm e}$, produced by the external constraints and introduced in Section~\ref{section_results_external}, remains as the main feature. Compared to the result from  the external constraints alone, in the combined analysis the EDGES data have the effect of reducing the probability at high $V_{\rm c}$. This is seen more clearly in the 1D $V_{\rm c}$ PDF, where we also notice that EDGES produces higher probabilities at lower $V_{\rm c}$; in particular, the region of low $V_{\rm c}$ contained within the $68\%$ probability volume is wider in the case of the combined constraints. From this PDF, we disfavor at $68\%$ confidence $V_{\rm c}<6.0$~km~s$^{-1}$ and $V_{\rm c}>46.6$~km~s$^{-1}$; this corresponds to $M_{\rm min}<3.9\times 10^6$~M$_\odot$ and $M_{\rm min}>1.8\times 10^9$~M$_\odot$ at $z=10$. We note that a range of values around the dip at $V_{\rm c}=16.5$~km~s$^{-1}$ (the atomic cooling threshold) is also outside the $68\%$ probability region. At $95\%$ confidence we derive the upper limit $V_{\rm c}=52$~km~s$^{-1}$, equivalent to $2.5\times 10^9$~M$_\odot$ at $z=10$.

The combined constraint on $\tau_{\rm e}$ is driven by \emph{Planck} + quasars + galaxies, with EDGES having a marginal contribution. Specifically, when we combine EDGES and the external observations we obtain the upper limit $\tau_{\rm e}<0.063$ at $68\%$ confidence, with the narrow range $\tau_{\rm e}=0.057-0.059$ outside the $68\%$ limits. This is almost identical to the result for \emph{Planck} + quasars + galaxies alone. It nonetheless reflects a broad consistency between EDGES and the external observations.

The shapes of the $f_*$, $f_{\rm X}$, and $\nu_{\rm min}$ 1D PDFs derived from the combined analysis are very close to those found using EDGES alone. As an example of the minor changes in the limits, the $68\%$ lower limit on $f_{\rm X}$ decreases from $0.0042$ for EDGES alone, to $0.0021$ in the combined analysis, which can be explained by the small decrease in the probability of high $f_{\rm X}$ produced by the external constraints on reionization. 

Finally, and as for $\tau_{\rm e}$, the combined constraint on $R_{\rm mfp}$ is mainly determined by the external observations. Comparing the 1D PDFs of $R_{\rm mfp}$ in Figure~\ref{figure_rmfp} we see that the PDF derived from the combined analysis (bottom row) is very similar to that obtained from the external observations alone (middle row). Although the EDGES data disfavor high values of $R_{\rm mfp}$ (top row), corresponding to sharper global signals, the combined analysis prefers high $R_{\rm mfp}$ and yields the lower limit $R_{\rm mfp}>27.5$~Mpc at $68\%$ confidence for fixed $\nu_{\text{min}}=0.5$~keV.

\capstartfalse
\begin{deluxetable}{lcccc}
%\tabletypesize{\scriptsize}
\tablewidth{0pt}
\tablecaption{Marginalized $68\%$ limits for $\tau_{\rm e}$ \label{table_taue}   }
\tablehead{Observation & Min & Max }
\startdata
quasars   $z=5.9$  \dotfill  & $0.068$ & $0.090$  \\
\\
quasar   $z=7.08$  \dotfill  & $0.055$ & $0.065$  \\
\\
quasar   $z=7.54$  \dotfill  & $0.055$ & $0.065$  \\
\\
galaxies $z=7$     \dotfill  & $0.055$ & $0.061$  \\
\\
\emph{Planck}      \dotfill  & $0.055$ & $0.063$  \\
\\
EDGES              \dotfill  & $0.055$ & $0.067$  \\
                     & $0.072$ & $0.080$  \\
\hline
\\
quasars            \dotfill  & $0.057$ & $0.067$  \\
\\
quasars + galaxies \dotfill  & $0.055$ & $0.057$  \\
                     & $0.059$ & $0.064$  \\
\\
quasars + galaxies + \emph{Planck} \dotfill  & $0.055$ & $0.058$  \\
                                     & $0.059$ & $0.063$  \\
\\
quasars + galaxies + \emph{Planck} + EDGES \dotfill & $0.055$ & $0.057$  \\ 
                                           & $0.059$ & $0.063$  
\enddata
\tablecomments{\\1) The $\tau_{\rm e}$ range explored is $0.055-0.090$. As the combined constraints prefer low $\tau_{\rm e}$, with high probability at $\approx 0.055$, we plan to extend the range below $0.055$ in future versions of \texttt{Global21cm}.\\2) These constraints are for $R_{\rm mfp}=30$~Mpc. The largest difference in the $\tau_{\rm e}$ limits when fixing instead $\nu_{\text{min}}=0.5$~keV occur when using EDGES data only. These two results are shown as cases A and B in Table~\ref{table_limits}. For the other observations or combinations, the difference in the $\tau_{\rm e}$ limits between $R_{\rm mfp}=30$~Mpc and $\nu_{\text{min}}=0.5$~keV is $\lesssim 10^{-3}$. \\3) In some cases, the $68\%$ probability volume is contained within two disjoint value ranges. These ranges are presented as two rows.}
\end{deluxetable}

\section{Discussion}
\label{section_discussion}

\subsection{Parameter Degeneracy}
\label{section_degeneracy}

As discussed in Section~\ref{section_results_external}, in our analysis, the parameters that drive the evolution of the neutral hydrogen fraction are $\tau_{\rm e}$ and $V_{\rm c}$, which suffer from a degeneracy in their effect on reionization.

Another example of degeneracy involves the parameters that have an effect on cosmic heating. The global $21$~cm signal is sensitive to the total energy injected into the gas, which depends on several quantities, such as the SFE and the SED of X-rays (including its shape and normalization). In our parametrization, the total number of X-ray photons that contribute to heating is determined by the values of $V_{\rm c}$, $f_*$, $f_{\rm X}$, $\nu_{\rm min}$, and $\alpha$. As a result, there are clear correlations between these parameters, which manifest themselves in the EDGES-only analysis (Figure~\ref{figure_vmin}) as diagonal trends on the 2D PDFs of $f_{\rm X}-V_{\rm c}$, $f_{\rm X}-f_*$, (and more weakly) $\nu_{\rm min}-V_{\rm c}$, $\nu_{\rm min}-f_{*}$, and $\nu_{\rm min}-f_{\rm X}$. The correlations demonstrate that, via X-ray heating, models with high $f_{\rm X}$ and high $V_{\rm c}$ have a similar signature within the EDGES band to cases with low $f_{\rm X}$ and low $V_{\rm c}$. The diagonal trend on the $f_{\rm X}-f_*$ 2D PDF reflects that, in the context of the global signal, the important parameter is $f_*f_{\rm X}$. This degeneracy is broken only for very low values of $f_*$, for which Ly-$\alpha$ coupling is inefficient. Finally, the degeneracy on the $\nu_{\rm min} - f_{\rm X}$ plane  reflects that hard spectra (high $\nu_{\rm min}$) generate less heating and require higher $f_{\rm X}$ to produce absorption troughs similar to models with low values of the two parameters.

The existing degeneracies in the global signal analysis arise because the parametrization of the 3D simulations used to train \texttt{Global21cm} is not optimized to represent the global signal alone, but instead to track the temporal evolution of the $21$~cm signal within a large cosmological volume from which higher order statistics, such as power spectra, can also be computed. In the future, an additional tool could be developed to establish consistency between the constraints obtained from radiometric and interferometric measurements, and reduce the degeneracies. Another important remaining task corresponds to finding a set of independent parameters to describe and constrain the astrophysics of the early Universe via the global $21$~cm signal alone. We leave this to future work.

\subsection{Comparison with Previous Results}

In this paper we constrain astrophysical processes during cosmic dawn and reionization, i.e., the same period constrained by \citet{greig2017a} and M18. However, as pointed out in Section~\ref{section_introduction}, in these works the astrophysical models were generated using the \texttt{21cmFAST} code, which differs from \texttt{Global21cm} in details of the processes modeled \citep[e.g.,][]{visbal2012, fialkov2013, cohen2016}. Other discrepancies include different parametrization  and  parameter ranges explored, as well as the parameters that were kept fixed during the exploration of the parameter space. These differences in modeling prevent us from making a quantitative comparison. The comparison is made even more difficult by the use of different external constraints. Therefore, here we limit ourselves to a high-level comparison only, leaving more detailed discussions for future work.

\citet{greig2017a} explored parameters relevant to reionization --- $\zeta$, $T_{\text{vir}}^{\text{min}}$, and $R_{\text{mfp}}$ --- assuming saturated X-ray heating and a star-formation efficiency of $5\%$. They found that the $\tau_{\rm e}$ estimate from \emph{Planck}~$2016$ and the $\bar{x}_{\rm H\scriptscriptstyle{I}}$ quasar constraints restrict the high-probability range in the joint $\zeta-T_{\text{vir}}^{\text{min}}$ PDF to a relatively narrow band across the plane (see the bottom row of their Figure $8$). Although this is a significant result, the band indicates a strong degeneracy between these parameters that prevents tight 1D marginalized constraints, in particular on $\zeta$. In M18 we found that combining the EDGES High-Band data with the \emph{Planck}~$2016$ + quasar constraints slightly decreased the degeneracy between $T_{\text{vir}}^{\text{min}}$ and $\zeta$ by reducing the probability of high $T_{\text{vir}}^{\text{min}}$ and high $\zeta$. We also incorporated into the analysis the X-ray heating parameters $L_{{\rm X}<2{\rm\;keV}}/{\rm SFR}$ and $E_0$, originally introduced in \citet{greig2017b}. The star-formation efficiency was still kept at $5\%$.  We obtained the following $68\%$ marginalized limits: (1) $5<\log_{10}\left(T_{\text{vir}}^{\text{min}}/{\rm K}\right)<5.6$, (2) $10<\zeta<148.4$, and (3) $0.62<E_0/{\rm keV}<1.5$. We also found that the $68\%$ confidence region of the soft-band X-ray luminosity was restricted to two ranges: $38<\log_{10}\left(L_{\text{X}<2~\rm keV}/{\rm SFR}\;{\rm /erg\;yr\;s}^{-1}\;{\rm M}^{-1}_{\odot}\right)<39$ and $40.8< \log_{10}\left(L_{\text{X}<2~\rm keV}/{\rm SFR}\;{\rm /erg\;yr\;s}^{-1}\;{\rm M}^{-1}_{\odot}\right)<42$.

The most direct comparison between the results of this paper and those in \citet{greig2017a} and M18, corresponds to the constraints on $V_{\rm c}$ reported here and their limits on $T_{\text{vir}}^{\text{min}}$. These two parameters directly depend on the mass of dark matter halos and are related via \citep{barkana2001}

% \footnote{In \cite{park2018} they introduce a new parametrization for 21cmFAST and 21CMMC that incorporates a model for the star formation efficiency that is a power-law function of the halo mass where the free parameters are a scaling parameter and the spectral index.}

%\begin{equation}
%T_{\text{vir}}^{\text{min}} = 1.98\times 10^4\left(\frac{\mu}{0.6}\right)\left(\frac{V_{\rm c}}{23.4}\right)^2,
%\end{equation}   

\begin{equation}
V_{\rm c} = 23.4 \sqrt{\left(\frac{0.6}{\mu}\right) \left(\frac{T_{\rm vir}^{\rm min}}{1.98\times 10^4}\right)}\;\;\;{\rm km~s}^{-1},
\end{equation}

\noindent where $\mu$ is the mean molecular weight, which varies between $0.59$ for a fully ionized and $1.22$ for neutral primordial gas. Qualitatively, the shape of the $V_{\rm c}$ and $T_{\text{vir}}^{\text{min}}$ 1D PDFs is similar: in the EDGES-only case the probability increases toward lower values, while in the combined analysis a peak occurs at intermediate values. Despite this qualitative agreement, quantitatively the constraints differ due to a number of discrepancies. The low-end cutoff in M18 is $T_{\rm vir}^{\rm min}=10^4$~K, which corresponds to $V_{\rm c} = 16.5$~km~s$^{-1}$ (assuming $\mu=0.59$). This means that molecular cooling halos are not accounted for in M18. Moreover, the large high-end cutoff in M18, $T_{\rm vir}^{\rm min}=10^6$~K, weights the probability towards higher values. This upper limit corresponds to $V_{\rm c} = 166$~km~s$^{-1}$, i.e, $M_{\rm min} \sim 8.2\times 10^{10}$~M$_{\odot}$ at $z=10$, while the upper limit in this paper is $76.5$~km~s$^{-1}$, corresponding to $M_{\rm min} \sim 8.0\times 10^{9}$~M$_{\odot}$ at $z=10$. The $95\%$ combined constraint on $T_{\text{vir}}^{\text{min}}$ derived in M18 implies $29.6< V_{\rm c}< 117.6$~km~s$^{-1}$, which corresponds to $4.6\times 10^8 < M_{\rm min} < 2.9\times 10^{10}$~M$_{\odot}$ at $z=10$; while our analysis here requires $V_{\rm c}< 52$~km~s$^{-1}$, or $M_{\rm min} < 2.5\times 10^9$~M$_{\odot}$ at $z=10$. Inclusion of small halos with $4.2<V_{\rm c}<16.5$~km~s$^{-1}$ in this work prevents us from determining a strong lower limit on $V_{\rm c}$. We note that taking $T_{\rm vir}^{\rm min}$ to be much higher than $10^5$~K should be disfavored by recent observations of high-$z$ galaxies of corresponding mass \citep[e.g.,][]{mashian2016, mirocha2017}.

A direct comparison between $V_{\rm c}$ in this work and $T_{\text{vir}}^{\text{min}}$ in M18 is not straightforward because M18 assumed a fixed $f_* = 5$\%, while here we vary $f_*$ over $0.1-50$\%. The effect of $f_*$ on the PDFs of $V_{\rm c}$ and $T_{\text{vir}}^{\text{min}}$ is as follows: for $f_*$ much lower than $5\%$, and all the other parameters fixed, $\zeta$ would be lower, making reionization slower and increasing the neutral fraction at a given redshift. Therefore, we would need to decrease $V_{\rm c} $ and $T_{\text{vir}}^{\text{min}}$ in order to produce more ionizing photons and compensate for the lower $\zeta$ when fitting to the $\bar{x}_{\rm H\scriptscriptstyle{I}}$ constraint at $z=5.9$. In other words, the $z=5.9$ constraint would prefer lower values of $V_{\rm c} $  and $T_{\text{vir}}^{\text{min}}$, shifting the peak of their 1D PDF to lower values, in broad agreement to what we see in our analysis when compared to M18. Increasing $f_*$ while leaving the other parameters fixed leads to a faster reionization and, thus, is not in conflict with the constraint at $z=5.9$.

% Thus, we should focus on comparing the probability peak at $V_{\rm c}>16.5$~km~s$^{-1}$. 

Comparing the constraints on the ionization parameters in \citet{greig2017a}, M18, and this work, is also non-trivial as $\tau_{\rm e}$, which we choose to constrain instead of $\zeta$, is an integrated quantity. In \citet{cohen2019} we checked that in \texttt{Global21cm}, when the {\it Planck} prior on $\tau_{\rm e}$ is applied, the allowed $\zeta$ is a growing function of $V_{\rm c}$ for $V_{\rm c}>16.5$~km~s$^{-1}$, in agreement with M18. For lower values of $V_{\rm c}$, $\zeta$ is nearly constant due to the effect of feedback mechanisms. In \texttt{Global21cm}, $\zeta$ is also a function of the SFE, which here we vary via $f_*$ while in M18 it is kept constant. Therefore, here we expect a larger scatter in $\zeta$ at a given  $V_{\rm c}$. Despite the differences in models and analyses, a high-level consistency between \citet{greig2017a}, M18, and this paper, is observed in the form of the narrow high-probability bands on the $\zeta-T_{\rm vir}^{\rm min}$ and $\tau_{\rm e}-V_{\rm c}$ 2D PDFs, which are obtained when applying the external $\tau_{\rm e}$ and $\bar{x}_{\rm H\scriptscriptstyle{I}}$ constraints. In addition, the middle and bottom rows of our Figure~\ref{figure_rmfp}, which show a preference for high $R_{\text{mfp}}$ by the external constraints, are in agreement with the equivalent result in \citet{greig2017a}.

The X-ray parameters from \texttt{21cmFAST} ($L_{\rm{X<2 keV}}/\text{SFR}$, $E_0$) and \texttt{Global21cm} ($f_\text{X}$, $\nu_{\rm min}$) are not very sensitive to the external $\tau_{\rm e}$ and $\bar{x}_{\rm H\scriptscriptstyle{I}}$ estimates and, thus, remain mainly constrained by the EDGES spectrum. Although in principle the two codes represent the same physical formalism, different assumptions and prior information lead to the exploration of different parameter ranges between M18 and here. Specifically, in M18 we explore scenarios where the IGM is heated by soft X-rays with (1) an X-ray spectral index $\alpha= -1$, (2) $E_0$ varying over $0.1-1.5$~keV, and (3) an X-ray luminosity equivalent to $f_{\rm X}$ in the range $\sim 3\times 10^{-2}-3\times 10^{2}$. In this work we thoroughly probe a wider range of heating scenarios by exploring $f_{\rm X}$ in the range $10^{-5}-10^{1}$ and $\nu_{\rm min}$ between $0.1$ and $3$~keV, assuming an X-ray spectral index $\alpha = -1.3$. A rough estimate, leaving aside the difference in the slope, low-energy cutoff, and X-ray energy range that goes into the definition of $L_{\rm X}$ ($E_0-10$~keV in M18 versus $\nu_{\rm min}-95$~keV here), shows that while in our models the combination $f_*f_{\rm X}$ varies between $10^{-8}$ and $5$, an equivalent combination in M18 (with $f_*=5\%$) varies between $1.6\times 10^{-4}$ and $1.6$. Although in this paper we do explore soft X-ray scenarios equivalent to those in M18, our broader parameter ranges enable us to probe many more `cold IGM' cases. Moreover, because here we probe models up to higher values of $\nu_{\rm min}$, and that they extend to higher X-ray energy (out to $95$~keV), on average our X-rays are harder and less efficient at heating than in M18. As a result, the global signals evaluated in this paper have, on average, deeper absorption troughs shifted to higher frequencies to which the EDGES High-Band spectrum is more sensitive. This enables us to derive the lower limits on $f_{\rm X}$ listed in Table~\ref{table_limits}.

Another aspect that leads to the differences in the X-ray constraints is the different range of $V_{\rm c}$ (equivalently, $T_{\rm vir}^{\rm min}$) explored in this paper ($4.2-76.5$~km~s$^{-1}$) and in M18 ($16.5-166$~km~s$^{-1}$). As the clearest example, Figure~$2$ of M18 shows regions of high probability for $T_{\rm vir}^{\rm min}\gtrsim10^{5.2}$~K that occur at low $L_{\text{X}<2~\rm keV}/{\rm SFR}$ and high $E_0$. Those high-probability regions are mostly outside the parameter space of this paper as the high-end $V_{\rm c}$ cutoff here is lower. Thus, they are not projected to the marginalized X-ray PDFs of this paper. However, for $T_{\rm vir}^{\rm min}\lesssim 10^{5.2}$~K, the $L_{\text{X}<2~\rm keV}/{\rm SFR}-T_{\rm vir}^{\rm min}$ and $E_0-T_{\rm vir}^{\rm min}$ PDFs in Figure~$2$ of M18 do resemble the equivalent $f_{\rm X}-V_{\rm c}$ and $\nu_{\rm min}-V_{\rm c}$ PDFs in Figure~\ref{figure_vmin} of this paper. This suggests that if a similar parameter space were explored, the constraints on parameters from \texttt{21cmFAST} and \texttt{Global21cm} would become more consistent.

%\footnote{However, the definition of $f_{\rm X}$ is not completely identical {\color{magenta}Note that the} because $L_{\rm X}$ in M18 is calculated at photon energies lower than $2$~keV while here we account for all the photons below $95$~keV. In addition, the heating history depends on the value of $f_*$, which in M18 is kept fixed at $5\%$.} 

% In this scenario, models with intermediate values of X-ray luminosity are disfavored because they correspond to global $21$~cm signals with absorption troughs that are relatively sharp and  distinguishable from the foregrounds and the uncertainty in the EDGES High-Band spectrum. 

\section{Summary}
\label{section_summary}

We report new constraints on high-$z$ astrophysical parameters derived from the EDGES High-Band measurement of the radio spectrum over $90-190$~MHz \citep{monsalve2017b}. We show that the spectrum is not only sensitive to reionization, i.e., the electron scattering optical depth and mean free path of ionizing photons, but can also constrain processes of star formation and heating during cosmic dawn. Specifically, we put limits on the minimum circular velocity (equivalent to the minimum mass) of star forming halos, the star formation efficiency, the X-ray efficiency of sources, and the low-energy cutoff of the X-ray SED. The definition and range of the parameters explored here correspond to the parametrization detailed  in \citet{cohen2017}. The models were generated using the new \texttt{Global21cm} interpolation tool \citep{cohen2019}. These models represent traditional physical scenarios and do not include the exotic physics proposed to explain the EDGES Low-Band measurement \citep{bowman2018}.

We compute the astrophysical parameter constraints within a Bayesian framework. First, we derive the constraints using the EDGES High-Band data alone. In this case the constraints depend on the sensitivity of the measurement --- limited by noise and systematic uncertainty --- to the spectral features of the $21$~cm signal within the range $90-190$~MHz, when simultaneously fitting a model that accounts for the foreground contribution. We then re-compute the constraints after incorporating into the analysis a prior on the electron scattering optical depth from \emph{Planck} \citep{planck2018} and estimates for $\bar{x}_{\rm H\scriptscriptstyle{I}}$ at $z\gtrsim 5.9$ from quasars \citep{mcgreer2015, greig2017c, banados2018} and Lyman Break galaxies \citep{mason2018}.

Using EDGES data alone, and after marginalization over the foreground parameters and the residual astrophysical parameters, we disfavor at $68\%$ confidence the following parameter ranges assuming a fixed $R_{\text{mfp}}=30$~Mpc:

\begin{enumerate}
\item High values of the minimum circular velocity of star-forming halos, $V_{\rm c} > 19.3$~km~s$^{-1}$. This value corresponds to a minimum halo mass of $1.3\times 10^{8}$~M$_{\odot}$ at $z=10$, which reflects that EDGES High-Band data are sensitive enough to constrain star formation in heavy halos. Lower values of $V_{\rm c}$ generate $21$~cm signals with absorption troughs at lower frequencies, which could be constrained more efficiently by Low-Band data. 
\item Intermediate values of star formation efficiency, $0.4\%< f_* < 3.9\%$. Low values of $f_*$ produce $21$~cm signals that fall in the High-Band range but have low amplitude, while high values of $f_*$ create troughs that are deep but wide. These types of signals cannot be disfavored with our current sensitivity.
\item Low values of the IGM X-ray heating efficiency, $f_\text{X} < 0.0042$. After exploring a wide dynamical range of cosmic heating, we robustly disfavor a `cold IGM' scenario.
\item High values of the electron scattering optical depth, $\tau_{\rm e}>0.08$, thus disfavoring early reionization. 
\item High values of the X-ray SED low-frequency cutoff, $\nu_{\text{min}}>2.3$~keV, constraining the X-ray hardness of the early sources.
\end{enumerate}
When fixing $\nu_{\text{min}}=0.5$~keV, the EDGES-only analysis also disfavors high values of the mean-free path of ionizing photons, $R_{\text{mfp}}>36.1$~Mpc.

Combining the EDGES High-Band data with the external observations primarily impacts the results for the parameters that most directly characterize the epoch of reionization: $\tau_{\rm e}$ and $R_{\text{mfp}}$. However, due to the dependence of the reionization history on star formation, the constraint on $\tau_{\rm e}$ is degenerate with $V_{\rm c}$, in particular for $V_{\rm c}>16.5$~km~s$^{-1}$, i.e., the atomic hydrogen cooling scenario.

In the combined analysis we obtain the optical depth upper limit $\tau_{\rm e}<0.063$ at $68\%$ confidence. We find a similar limit, $\tau_{\rm e}<0.064$, using only the neutral fraction estimates from quasars and LBGs. This reflects a broad agreement between independent observations despite the different models used for the redshift evolution of the neutral fraction. The EDGES contribution to the combined $\tau_{\rm e}$ constraint is marginal.

For $V_{\rm c}$, the combined analysis disfavors at $68\%$ confidence the ranges $V_{\rm c}<6.0$~km~s$^{-1}$ and $V_{\rm c}>46.6$~km~s$^{-1}$, while at $95\%$ it rules out $V_{\rm c}>52.0$~km~s$^{-1}$. This result indicates that EDGES High-Band + \emph{Planck} + quasars + galaxies require the existence of halos with minimum cooling mass below $2.5\times 10^9$~M$_\odot$ at $z=10$. Interestingly, this is consistent with the EDGES absorption feature reported in Low-Band data, which requires efficient star formation in halos well below $10^{10}$~M$_{\odot}$ \citep{mirocha2019}.

The combined analysis reverses the shape of the $R_{\text{mfp}}$ PDF relative to the result from EDGES data alone, and assigns higher probabilities to higher values of this parameter, i.e., scenarios with faster growth of ionized bubbles and, therefore, faster reionization. Specifically, at $68\%$ confidence we obtain the lower limit $R_{\text{mfp}}>27.5$~Mpc. Faster reionization scenarios are more compatible with the $\bar{x}_{\rm H\scriptscriptstyle{I}}$ upper limit at $z=5.9$ from \citet{mcgreer2015} combined with the reports of ongoing reionization at $z\gtrsim 7$ and a low optical depth.

Finally, the external observations do not impact significantly the results for the other astrophysical parameters --- $f_*$, $f_{\rm X}$, and $\nu_{\rm min}$---, which remain mainly constrained by the EDGES High-Band spectrum.

The results of this paper are in broad agreement with the analyses of \citet{greig2017a} and \citet{monsalve2018}, which explored astrophysical models generated with the \texttt{21cmFAST} code \citep{mesinger2007, mesinger2011}. Nonetheless, noticeable differences occur with the results for $V_{\rm c}$ and the X-ray heating parameters in \citet{monsalve2018}. These discrepancies are primarily due to: (1) keeping in this paper $f_*$ as a free parameter instead of fixing it at $5\%$ as in \citet{monsalve2018}; (2) exploring here a different range for $V_{\rm c}$, which considers star formation in both, atomic and molecular cooling halos; and (3) exploring here wider ranges for $f_{\rm X}$ and $\nu_{\rm min}$, which extend to scenarios of very inefficient heating due to weak or hard X-ray sources.

We leave for future work detailed comparisons with results for models from \texttt{21cmFAST}, as well as analyses that incorporate measurements from EDGES Low-Band, which should increase the sensitivity to $21$~cm signals whose main features lie below $\sim100$~MHz.

% Considering this limit, as well as the trend of decreasing $\tau_{\rm e}$ in recent \emph{Planck} releases, we recognize the need to extend the $\tau_{\rm e}$ range accessible through \texttt{Global21cm} to values below the current low-end cutoff of $\tau_{\rm e}=0.055$, in future versions of the code.

%We leave for future work detailed comparisons with results for models from \texttt{21cmFAST}. 

%Even though we do not use EDGES Low-Band data in this paper, the results presented here are consistent with the detected signal. Specifically, non-detection in High-Band data implies that the minimum cooling mass of star-forming halos should be below {\color{red}$8.3\times 10^{7}$ M$_{\odot}$ (68\%) WHERE DOES THIS NUMBER COME FROM??? DESCRIBE OR ELABORATE MORE, OR FIX IT IF NECESSARY}, while detection of the signal in the Low-Band spectrum requires $M_{\rm min} \lesssim {\rm few} \times 10^{8}$~M$_{\odot}$ \citep[][and work in prep.]{mirocha2018}.

\acknowledgements
We are grateful to Bradley Greig, Eduardo Ba\~nados, and Charlotte Mason for providing to us the hydrogen neutral fraction PDFs derived from quasar ULASJ1120+0641, quasar ULASJ1342+0928, and high-$z$ galaxies, respectively. We also thank Bradley Greig, Kohei Inayoshi, Nicholas Kern, Andrei Mesinger, and Jordan Mirocha for useful discussions. This work was supported by the NSF through research awards for the Experiment to Detect the Global EoR Signature (AST-0905990, AST-1207761, and AST-1609450). R.A.M. was supported by the NASA Solar System Exploration Virtual Institute cooperative agreement 80ARC017M0006, and by the NASA Ames Research Center grant NNX16AF59G. R.A.M. conducted part of this work at the Astrophysics and Cosmology Research Unit, University of KwaZulu-Natal, South Africa. A.F. is supported by the Royal Society University Research Fellowship. For R.B. and A.C., this publication was made possible by the ISF-NSFC joint research program (grant No. 2580/17) and through the support of a grant from the John Templeton Foundation; the opinions expressed in this publication are those of the authors and do not necessarily reflect the views of the John Templeton Foundation. Computations in this paper were run on the {\it Odyssey} cluster supported by the FAS Division of Science, Research Computing Group at Harvard University. EDGES is located at the Murchison Radio-astronomy Observatory. We acknowledge the Wajarri Yamatji people as the traditional owners of the Observatory site. We thank CSIRO for providing site infrastructure and support.

\emph{Software}: Ipython (\url{http://dx.doi.org/10.1109/MCSE.2007.53}), Numpy (\url{http://dx.doi.org/10.1109/MCSE.2011.37}), Scipy (\url{https://doi.org/10.5281/zenodo.1036423}), Matplotlib (\url{https://doi.org/10.5281/zenodo.573577}), Astropy \citep{astropy2013}, Healpy \citep{gorski2005}, h5py (\url{https://doi.org/10.5281/zenodo.877338}).

\section*{ORCID \lowercase{i}D\lowercase{s}}
\noindent Raul A. Monsalve \href{https://orcid.org/0000-0002-3287-2327}{https://orcid.org/0000-0002-3287-2327} \\
Anastasia Fialkov \href{https://orcid.org/0000-0002-1369-633X}{https://orcid.org/0000-0002-1369-633X} \\
Judd D. Bowman \href{https://orcid.org/0000-0002-8475-2036}{https://orcid.org/0000-0002-8475-2036} \\
Alan E. E. Rogers \href{https://orcid.org/0000-0003-1941-7458}{https://orcid.org/0000-0003-1941-7458} \\
Thomas J. Mozdzen \href{https://orcid.org/0000-0003-4689-4997}{https://orcid.org/0000-0003-4689-4997} \\
Nivedita Mahesh \href{https://orcid.org/0000-0003-2560-8023}{https://orcid.org/0000-0003-2560-8023}

\clearpage

\newpage

\end{document}